\documentclass[useAMS,usenatbib, usegraphicx, referee]{mn2e}

\title[Can stellar mass black holes be quark stars?]{Can stellar mass black holes be quark stars?}
\author[Kov\'{a}cs, Cheng and Harko]{Z. Kov\'{a}cs$^{1,2,3}$\thanks{E-mail:
zkovacs@mpifr-bonn.mpg.de}, K. S. Cheng$^{1}$\thanks{E-mail:
hrspksc@hkucc.hku.hk} and T. Harko$^{1}$\thanks{E-mail:
harko@hkucc.hku.hk}
\\
$^{1}$Department of Physics and Center for Theoretical and Computational Physics, The University of Hong Kong, Hong Kong, P. R. China\\
$^{2}$Max-Planck-Institut f\"{u}r Radioastronomie, Auf dem H\"{u}gel 69,
53121 Bonn, Germany\\
$^{3}$Department of Experimental Physics, University of Szeged, D\'{o}m T%
\'{e}r 9, Szeged 6720, Hungary}
\begin{document}


\pagerange{\pageref{firstpage}--\pageref{lastpage}} \pubyear{2002}

\maketitle

\label{firstpage}

\begin{abstract}
We investigate the possibility that stellar mass black holes, with masses in the range of $3.8M_{\odot}$ and $6M_{\odot}$, respectively, could be in fact quark stars in the Color-Flavor-Locked (CFL) phase. Depending on the value of the gap parameter, rapidly rotating CFL quark stars can achieve much higher masses than standard neutron stars, thus making them possible stellar mass black hole candidates. Moreover, quark stars have a very low luminosity and a completely absorbing surface - the infalling matter on the surface of the quark star is converted into quark matter. A possibility of distinguishing CFL quark stars from stellar mass black holes could be through the study of thin accretion disks around rapidly rotating quark stars and Kerr type black holes, respectively. Furthermore, we show that the radiation properties of accretion disks around black holes and CFL quark stars are also very similar. However, strange stars exhibit a low luminosity, but high temperature bremsstrahlung spectrum, which, in combination with the emission properties of the accretion disk, may be the key signature to differentiate massive strange stars from black hole.

\end{abstract}


\section{Introduction}

Current physics suggests that compact objects with mass functions larger than 3-4 solar masses must be black holes \citep{Ru74}. From a theoretical point of view, black holes have been considered, in the framework of general relativity, for almost one hundred years. However, their first observational evidence was not found until quite recently, during the past three decades. In the late 1960s X-ray detectors onboard satellites revolutionized astronomy with the discovery of a large population of luminous X-ray sources in the Galaxy. Optical counterparts associated with high mass X-ray binaries (HMXBs) have also been identified, and one of the first to be identified was the supergiant star HD 226868, associated with HMXB Cyg X-1. The determination of the mass function equation $f\left(M_x\right)$, which relates the mass of the compact object $M_x$ with that of the companion star $M_c$ and to the inclination angle $i$ indicated that $M_x>4M_{\odot}$ \citep{Bo75}. In 1975 the satellite Ariel V detected the X-ray source A-0620-00, which belongs to a class of X-ray transients (XRT). The mass function of the compact star is $3.2\pm0.2M_{\odot}$, which slightly exceeds the maximum mass allowed for a stable neutron star \citep{ClRe86}.  In 1989, the X-ray satellite Ginga discovered the XRT V404 Cyg. The mass function for this system implies the presence of a compact object with mass greater than $6M_{\odot}$. Since then, many other stellar mass black hole candidates have been found, with seven of them having mass functions in excess of $5M_{\odot}$. Presently, there are around 20 known stellar mass black holes, like, for example, XTE J1650-500, with a mass of 3.8 solar masses \citep{Orosz}, GRO J0422+32 with a mass between 3-5 solar masses \citep{Gel}, GRO J1655-40 with a mass of around 6 solar masses \citep{Udal} etc.  However, it is estimated that in the Milky Way alone there should be at least 1000 dormant black hole XRTs, while the total number of stellar mass black holes (isolated and in binaries) could be as large as 100 million \citep{Ro98}. The stellar mass black hole holes have been observed in close binary systems, in which transfer of matter from a companion star to the black hole occurs. The energy released in the fall heats  up the matter to temperatures of several hundred million degrees, and it is radiated in X-rays. The black hole therefore is observable in X-rays, whereas the companion star can be observed with optical telescopes. The energy released for black holes and neutron stars is of the same order of magnitude.  Therefore, on the one hand,  black holes and neutron stars are often difficult to distinguish. On the other hand, since many types of compact objects formed from so called exotic matter (boson stars, axion stars, condensed stars etc.) have properties similar to the standard black holes \citep{To02,YuNaRe04}, distinguishing between all these classes of compact object is a fundamental challenge for contemporary astronomy and astrophysics. There are also important observational evidences for massive black holes, like the intermediate-mass black holes (which can be found in the center of globular clusters) and the supermassive black holes in the center of the Milky Way and of the active galaxies \citep{Ho}.

The quark structure of the nucleons, suggested by quantum chromodynamics,
indicates the possibility of a hadron-quark phase transition at high densities
and/or temperatures \citep{It70, Bo71,Wi84}.  If the hypothesis of the quark matter is true,
then some of neutron stars could actually be quark stars, built entirely
of quark matter \citep{Al86,Ha86}. For a general review of quark star
properties see \citet{Ch98}.

There are several proposed mechanisms for the formation of quark stars.
Quark stars are expected to form during the collapse of the core of a
massive star, after the supernova explosion, as a result of a first or second
order phase transition, resulting in deconfined quark matter \citep{Da}. The
proto-neutron star core or the neutron star core is a favorable environment
for the conversion of ordinary matter to strange quark matter \citep{ChDa, Chan09}.
Another possibility is that some neutron stars in low-mass X-ray binaries
can accrete sufficient mass to undergo a phase transition to become strange
stars \citep{Ch96}. This mechanism has also been proposed as a source of
radiation emission for cosmological $\gamma $-ray bursts \citep{Ch98a}.

Most of the investigations of quark star properties have been done
within the framework of the so-called MIT bag model, with the
energy density $\rho c^{2}$ and pressure $p$ of a quark-gluon
plasma related by the equation of state (EOS) $p=(\rho -4B)c^{2}/3$, where $B$ is the
difference between the energy density of the perturbative and
non-perturbative QCD vacuum (the bag constant). For different values of
the bag constant and for different equations of state of the
quark matter a
complete description of the basic astrophysical properties (mass,
radius, eccentricity, Keplerian frequency etc.) of both static and
rotating quark stars has been obtained \citep{Wi84, Ha86, Al86, De98, Go00, HaCh02}. Quark stars can reach much shorter periods than
neutron stars, of the order of $0.5$ ms. Quark stars could have a radius significantly less than that of neutron stars \citep{Ch98}. $r$-mode instabilities in rapidly rotating quark stars lead to specific signatures in the evolution of pulsars with periods below $2.5$ ms.
 If quark matter is absolutely stable, some pulsars, in which the conditions for a hadron-quark phase transition are realized,  could consist of quark matter. Some data on pulsar properties are consistent with this assumption \citep{Ma00}.



The existence of a large variety of color superconducting phases of quark matter
at high densities has been suggested and intensively investigated \citep{cfl1, 3a, 3c, cfl3}. At
ultra-high density, matter is expected to form a degenerate Fermi
gas of quarks in which the Cooper pairs of quarks with very high
binding energy condensate near the Fermi surface (color
superconductor). Such a state is significantly more bound than
ordinary quark matter. It is now widely accepted that at extremely
high density the ground state of quark matter is the
superconducting Color-Flavor-Locked (CFL) phase, and that this
phase of matter rather than nuclear matter may be the ground state
of hadronic matter \citep{cfl3}. The existence of the CFL phase can
enhance the possibility of the existence of a pure stable quark
star. The maximum mass of the static quark stars in the CFL model is given by $M_{\max}=1.96M_{\odot}\left(1+\delta \right)/\sqrt{B_{60}}$, where $B_{60}=B\;{\rm MeV}\;{\rm fm}^{-3})/60\;{\rm MeV}\;{\rm fm}^{-3}$, and $\delta =0.15\left(\Delta/100\; {\rm MeV}\right)^2\left(60\;{\rm MeV}\;{\rm fm}^{-3}/B\right)$, with $\Delta $ the gap energy \citep{HoLu04}. Hence large maximum masses ($M>4M_{\odot}$) can be found for standard values of $B$ and $\Delta \geq 250$ MeV. Rotation can also significantly increase the mass of the CFL quark stars. For a discussion of the maximum mass of both rotating and non-rotating general relativistic objects see \cite{Sterev}. For rotating stars the maximum mass can be of the order of $M_{max}\approx 6.1M_{\odot}\left(2\times 10^{14}\;{\rm g}\;{\rm cm}^{-3}/\rho _m\right)^{1/2}$, where $\rho _m$ is the matching density.

Recent observations suggest that around
most of the active galactic nuclei (AGN's) or black hole candidates there
exist gas clouds surrounding the central compact object, and an associated
accretion disc, on a variety of scales from a tenth of a parsec to a few
hundred parsecs \citep{UrPa95}. These clouds are assumed to form a
geometrically and optically thick torus (or warped disc), which absorbs most
of the ultraviolet radiation and the soft X-rays. The gas exists in either
the molecular or the atomic phase.  Hence,
important astrophysical information can be obtained from the observation of
the motion of the gas streams in the gravitational field of compact objects.

The determination of the accretion rate for an astrophysical object can give
a strong evidence for the existence of a surface of the object. A model in
which Sgr A*, the $3.7\times 10^6 M_{\odot }$ super massive black hole
candidate at the Galactic center, may be a compact object with a thermally
emitting surface was considered in \cite{BrNa06}.  Given the very low quiescent
luminosity of Sgr A* in the near-infrared, the existence of a hard surface,
even in the limit in which the radius approaches the horizon, places a
severe constraint on the steady mass accretion rate onto the source, ${\dot M%
}\le 10^{-12} M_{\odot}$ yr$^{-1}$. This limit is well below the minimum
accretion rate needed to power the observed submillimeter luminosity of Sgr
A*, ${\dot M}\ge 10^{-10} M_{\odot}$ yr$^{}$.
Thus, from the determination of the accretion rate it follows that Sgr A*
does not have a surface, that is, it must have an event horizon. Therefore
the study of the accretion processes by compact objects is a powerful
indicator of their physical nature.

The first comprehensive theory of accretion disks around compact general relativistic objects was
constructed in \citet{ShSu73}. This theory was extended to the general
relativistic models of the mass accretion onto rotating black holes in \citet%
{NoTh73}. These pioneering works developed thin steady-state accretion
disks, where the accreting matter moves in Keplerian orbits. The
hydrodynamical equilibrium in the disk is maintained by an efficient cooling
mechanism via radiation transport. The photon flux emitted by the disk
surface was studied under the assumption that the disk emits a black body
radiation. The properties of radiant energy flux over the thin accretion
disks were further analyzed in \citet{PaTh74} and in \citet{Th74}, where the
effects of the photon capture by the hole on the spin evolution were
presented as well. In these works the efficiency with which black holes
convert rest mass into outgoing radiation in the accretion process was also
computed.

The emissivity properties of the accretion disks have been recently investigated for different classes of compact objects, such as rotating and non-rotating boson or fermion stars \citep{To02,YuNaRe04}, as well as for the modified $f(R)$ type theories of gravity \citep{mod}, for brane world black holes \citep{mod1}, and for wormholes \citep{mod2, mod3}. The radiation power per unit area, the temperature of the disk and the spectrum of the emitted radiation were given, and compared with the case of a Schwarzschild or Kerr-Newman black holes of an equal mass.

Since quark stars in the CFL phase can have higher maximum masses than ordinary neutron stars \citep{HoLu04}, which, in the case of rapidly rotating stars, can reach values as high as $6-7M_{\odot}$, the possibility that some stellar mass black holes are actually quark stars cannot be excluded {\it a priori}.  The CFL quark stars, being more massive than the presumed neutron-star limit of $3-4M_{\odot}$, can describe compact objects that do not fit into the framework of standard neutron star models. It is the purpose of the present paper to propose an observational test of this hypothesis, by considering a comparative study of the properties of the thin accretion disks around rapidly rotating CFL quark stars and black holes, respectively. In particular, we consider a comparative study of the basic physical parameters describing the disks, like the emitted energy flux, the temperature distribution on the surface of the disk, as well as the spectrum of the equilibrium radiation. Due to the differences in the exterior geometry, the thermodynamic and electromagnetic properties of the disks (energy flux, temperature distribution and equilibrium radiation spectrum) are different for these two classes of compact objects, thus giving clear observational signatures, which may allow to distinguish CFL quark stars from black holes.

The present paper is organized as follows.  The properties of the general relativistic thin accretion disks onto compact objects are briefly described in Section II. In Section III we present the metric properties for Kerr black holes as well as for the CFL quark stars. In Section IV we consider the radiation
flux, spectrum and efficiency of thin accretion disks onto Kerr black holes and quark stars. We discuss and conclude our results in Section V.

\section{Thin accretion disks onto general relativistic compact objects}

 Accretion discs are flattened astronomical objects made of rapidly rotating gas which slowly spirals onto a central gravitating body, with its gravitational energy degraded to heat. A fraction of the heat converts into
radiation, which partially escapes, and cools down the accretion disc.
The only information that we have about accretion disk physics comes from
this radiation, when it reaches radio, optical and $X$-ray telescopes,
allowing astronomers to analyze its electromagnetic spectrum, and its time
variability. The efficient cooling via the radiation over the disk surface prevents the disk
from cumulating the heat generated by stresses and dynamical
friction. In turn, this equilibrium causes the disk to stabilize
its thin vertical size. The thin disk has an inner edge at the
marginally stable orbit of the compact object potential, and the
accreting plasma has a Keplerian motion in higher orbits.

For the general relativistic case the theory of mass accretion around rotating black holes was developed  by \citet{NoTh73}. They
extended the steady-state thin disk models introduced by
\citet{ShSu73} to the case of the curved space-times, by adopting the equatorial
approximation for the stationary and axisymmetric geometry. The time- and
space-like Killing vector fields $(\partial / \partial t)^{\mu}$ and $%
(\partial / \partial \phi)^{\mu}$ describe the symmetry properties of this
type of space-time, where $t$ and $r$ are the Boyer-Lindquist time and
radial coordinates, respectively.
A thin accretion disk is an accretion disk such that in
cylindrical coordinates $(r,\phi ,z)$ most of the
matter lies close to the radial plane.

For the thin accretion disk its vertical size (defined along the $z$-axis)
is negligible, as compared to its horizontal extension (defined along the
radial direction $r$), i.e, the disk height $H$, equal to the maximum
half thickness of the disk, is always much smaller than the characteristic radius $R$ of
the disk, $H \ll R$. The thin disk is in hydrodynamical equilibrium, and the pressure gradient
and a vertical entropy gradient in the accreting matter are negligible.
In the steady-state accretion disk models, the mass accretion rate $\dot{M%
}_{0} $ is supposed to be constant in time, and the physical quantities of
the accreting matter are averaged over a characteristic time scale, e.g. $%
\Delta t$, and over the azimuthal angle $\Delta \phi =2\pi $, for a total period of
the orbits and for the height $H$. The plasma moves in Keplerian orbits around
the compact object, with a rotational velocity $\Omega $, and the plasma
particles have a specific energy $\widetilde{E}$, and specific angular
momentum $\widetilde{L}$, which depend only on the radii of the orbits. The
particles are orbiting with the four-velocity $u^{\mu }$ in a disk having an
averaged surface density $\Sigma $. The accreting matter is modeled by an
anisotropic fluid source, where the rest mass density $\rho _{0}$ (the specific internal energy
is neglected), the energy flow vector $q^{\mu }$ and the stress tensor $%
t^{\mu \nu }$ are measured in the averaged rest-frame. The energy-momentum
tensor describing this source takes the form
\begin{equation}
T^{\mu \nu }=\rho _{0}u^{\mu }u^{\nu }+2u^{(\mu }q^{\nu )}+t^{\mu \nu }\;,
\end{equation}
where $u_{\mu }q^{\mu }=0$, $u_{\mu }t^{\mu \nu }=0$. The four-vectors of
the energy and of the angular momentum flux are defined by
$-E^{\mu }\equiv T_{{}}^{\mu }{}_{\nu }(\partial /\partial t)^{\nu }$ and $
J^{\mu }\equiv T_{{}}^{\mu }{}_{\nu }(\partial /\partial \phi )^{\nu }$,
respectively. The four dimensional conservation laws
of the rest mass, of the energy and of the angular momentum of the plasma
provide the structure equations of the thin disk. By integrating the
equation of the rest mass conservation, $\nabla _{\mu }(\rho _{0}u^{\mu })=0$, %
 it follows that the time averaged
 accretion rate  $\dot{M_{0}}$ is independent of the disk radius:
$\dot{M_{0}}\equiv -2\pi r\Sigma u^{r}=\mbox{const}$,
where a dot represents the derivative with respect to the time coordinate \citep{PaTh74}.
The averaged rest mass density  is defined by
$\Sigma (r)=\int_{-H}^{H}\langle \rho _{0}\rangle dz$,
where $\langle \rho _{0}\rangle $ is the rest mass density averaged  over $\Delta
t$ and $2\pi $. The conservation law $\nabla _{\mu }E^{\mu }=0$ of the
energy can be written in an integral form as
\begin{equation}
\lbrack \dot{M}_{0}\widetilde{E}-2\pi r\Omega W_{\phi }{}^{r}]_{,r}=4\pi
\sqrt{-g}F\widetilde{E}\;\;,  \label{conslawofE}
\end{equation}%
where a comma denotes the derivative with respect to the radial coordinate $r$. Eq.~(\ref{conslawofE}) shows the balance between the energy transported by the rest mass flow, the dynamical stresses in the disk, and the energy radiated away
from the surface of the disk, respectively. The torque $W_{\phi }{}^{r}$ in Eq.~(\ref{conslawofE}) is given by
$W_{\phi }{}^{r}=\int_{-H}^{H}\langle t_{\phi }{}^{r}\rangle dz$,
where $\langle t_{\phi }{}^{r}\rangle $ is the $\phi -r$ component of the stress
tensor, averaged over $\Delta t$ and over a $2\pi $ angle. The law of the angular momentum
conservation, $\nabla _{\mu }J^{\mu }=0$,  states in its integral form the
balance of the three forms of the angular momentum transport,
\begin{equation}
\lbrack \dot{M}_{0}\widetilde{L}-2\pi rW_{\phi }{}^{r}]_{,r}=4\pi \sqrt{-g}F%
\widetilde{L}\;\;.  \label{conslawofL}
\end{equation}

By eliminating $W_{\phi}{}^{r}$ from Eqs. (\ref{conslawofE}) and (\ref%
{conslawofL}), and by applying the universal energy-angular momentum relation
$dE=\Omega dJ$ for circular geodesic orbits in the form $\widetilde{E}%
_{,r}=\Omega\widetilde{L}_{,r}$, the flux of the
radiant energy over the disk can be expressed in terms of the specific energy, angular
momentum and the angular velocity of the black hole. Then the flux integral
leads to the expression of the energy flux $F(r)$, which is given by \citep{PaTh74, Th74}
\begin{equation}
F(r)=-\frac{\dot{M}_0}{4\pi\sqrt{-g}} \frac{\Omega_{,r}}{(\widetilde{E}%
-\Omega\widetilde{L})^{2}} \int_{r_{ms}}^{r}(\widetilde{E}-\Omega\widetilde{L%
})\widetilde{L}_{,r}dr\;,  \label{F}
\end{equation}
where the no-torque inner boundary conditions were also prescribed \citep%
{PaTh74}. This means that the torque vanishes at the inner edge of the disk,
since the matter at the marginally stable orbit $r_{ms}$ falls
freely into the black hole, and cannot exert considerable torque on the
disk. The latter assumption is valid as long as strong magnetic fields do
not exist in the plunging region, where matter falls into the hole.

The presence of magnetic fields would give a non zero torque at the inner edge
of the disk. If the central object has a high magnetic field, the disk will be truncated at a radius larger than that of the last stable orbit for an unmagnetized star. The presence of high magnetic fields will also have a significant effect on the energy and angular momentum transport between the rotating central object and the accretion disk.
A central object surrounded by a magnetosphere can also exert a torque on the accretion disk, via the closed magnetic flux lines, connecting the disk and the rotating center. Then, the latter injects angular momentum into the orbiting plasma, which in turn increases the flux radiated by the disk surface  \citep{Li02}. In magnetized thin disks, the magnetic energy and magnetic stresses give also a contribution to the radiated flux.
The gravitational potential is determined by the nature of the rotating central object, whereas the presence of the magnetosphere involves various magnetohydrodynamic (MHD) effects, complicating the picture.
Since the magnetic torque is proportional to the difference in the angular velocities of the disk and of the magnetic field \citep{Li02},
and the contribution of the magnetic energy and stresses to the disk radiation has the same proportionality factor for magnetized thin accretion disks \citep{KBG},
this simplification is valid as long as the plasma co-rotates with the magnetosphere.
Numerical models of MHD accretion flows indeed show that the rotating plasma can be coupled to the magnetic field in turbulent magnetized disks where the plasma particles cannot slip along the field lines \citep{McNa07}. Assuming this type of turbulent accretion we remain in the framework of the hydrodynamical description. In the present study we neglect all the effects related to magnetic fields, by assuming that the magnetic fields in the central object as well as in the disk are very low, and we consider only the pure hydrodynamical description of the disk accretion, that is, we focus only on the effects of gravity.

Once the geometry of the space-time is known, we can derive the time
averaged radial distribution of photon emission for accretion disks around
black holes, and determine the efficiency of conversion of the rest mass into
outgoing radiation. After obtaining the radial dependence of
the angular velocity $\Omega $, of the specific energy $\widetilde{E}$ and
of the specific angular momentum $\widetilde{L}$ of the particles moving on
circular orbits around the black holes, respectively, we can compute the
flux integral given by Eq.~(\ref{F}).

Let us consider an arbitrary stationary and axially symmetric geometry,
\begin{equation}
ds^{2}=g_{tt}dt^{2}+g_{t\phi }dtd\phi +g_{rr}dr^{2}+g_{\theta \theta
}d\theta ^{2}+g_{\phi \phi }d\phi ^{2}\;,  \label{ds2rcoappr}
\end{equation}%
where in the equatorial approximation ($|\theta -\pi /2|\ll 1$) the metric functions $g_{tt}$, $g_{t\phi }$, $g_{rr}$, $g_{\theta
\theta }$ and $g_{\phi \phi }$ depend only on the radial coordinate $r$. The geodesic equations take the form
\begin{equation}
 \frac{dt}{d\tau }=\frac{\widetilde{E}g_{\phi \phi }+%
\widetilde{L}g_{t\phi }}{g_{t\phi }^{2}-g_{tt}g_{\phi \phi }},
\; \frac{d\phi }{d\tau }=-\frac{\widetilde{E}g_{t\phi }+%
\widetilde{L}g_{tt}}{g_{t\phi }^{2}-g_{tt}g_{\phi \phi }},
\end{equation}%
and
\begin{equation}\label{geodeqs3}
g_{rr}\left( \frac{dr}{d\tau }\right) ^{2}=V(r),
\end{equation}%
respectively, where $\tau $ is the affine parameter, and the potential term $V(r)$ is defined by
\begin{equation}
V(r)\equiv -1+\frac{\widetilde{E}^{2}g_{\phi \phi }+2\widetilde{E}%
\widetilde{L}g_{t\phi }+\widetilde{L}^{2}g_{tt\texttt{}}}{g_{t\phi
}^{2}-g_{tt}g_{\phi \phi }}\;.
\end{equation}

For circular orbits in the equatorial plane the conditions  $V(r)=0$ and $V_{,r}(r)=0$, respectively, must
hold. These conditions give the specific energy $\widetilde{E}$, the specific angular
momentum $\widetilde{L}$ and the angular velocity $\Omega $ of particles
moving on circular orbits around spinning general relativistic stars as%
\begin{eqnarray}
\widetilde{E} &=&-\frac{g_{tt}+g_{t\phi }\Omega }{\sqrt{-g_{tt}-2g_{t\phi
}\Omega -g_{\phi \phi }\Omega ^{2}}}\;,  \label{tildeE} \\
\widetilde{L} &=&\frac{g_{t\phi }+g_{\phi \phi }\Omega }{\sqrt{%
-g_{tt}-2g_{t\phi }\Omega -g_{\phi \phi }\Omega ^{2}}},  \label{tildeL} \\
\Omega  &=&\frac{d\phi }{dt}=\frac{-g_{t\phi ,r}+\sqrt{(g_{t\phi
,r})^{2}-g_{tt,r}g_{\phi \phi ,r}}}{g_{\phi \phi ,r}}\;.\label{Omega}
\end{eqnarray}
The marginally stable orbits $r_{ms}$ around the central object are determined by
the condition $\left.V_{,rr}(r)\right|_{r=r_{ms}}=0$, which gives
\begin{equation}
\left[\widetilde{E}^{2}g_{\phi \phi ,rr}+2\widetilde{E}\widetilde{L}g_{t\phi ,rr}+%
\widetilde{L}^{2}g_{tt ,rr}-(g_{t\phi }^{2}-g_{tt}g_{\phi \phi
})_{,rr}\right]_{r=r_{ms}}=0\;.  \label{stable}
\end{equation}

By inserting Eqs.~(\ref{tildeE})-(\ref{tildeL}) into Eq.~(\ref{stable}), and
solving the resulting equation for $r_{ms}$, we obtain the marginally stable
orbits, once the metric coefficients $g_{tt}$, $g_{t\phi }$ and $%
g_{\phi \phi }$ are explicitly given.



The accreting matter in the steady-state thin disk model is supposed to be
in thermodynamical equilibrium. Therefore the radiation emitted by the disk
surface can be considered as a perfect black body radiation, where the
energy flux is given by $F(r)=\sigma T^{4}(r)$ ($\sigma $ is the
Stefan-Boltzmann constant), and the observed luminosity $L\left( \nu \right) $ has a redshifted black body spectrum \citep{To02}:
\begin{equation}
L\left( \nu \right) =4\pi d^{2}I\left( \nu \right) =\frac{8}{\pi c^2 }\cos \gamma \int_{r_{i}}^{r_{f}}\int_0^{2\pi}\frac{\nu^{3}_e r d\phi dr }{\exp \left( h\nu_e/T\right) -1}.
\end{equation}

Here $d$ is the distance to the source, $I(\nu )$ is the Planck
distribution function, $\gamma $ is the disk inclination angle, and $r_{i}$
and $r_{f}$ indicate the position of the inner and outer edge of the disk,
respectively. We take $r_{i}=r_{ms}$ and $r_{f}\rightarrow \infty $, since
we expect the flux over the disk surface vanishes at $r\rightarrow \infty $
for any kind of general relativistic compact object geometry. The emitted frequency is given by $\nu_e=\nu(1+z)$, where the redshift factor can be written as
\begin{equation}
1+z=\frac{1+\Omega r \sin \phi \sin \gamma }{\sqrt{ -g_{tt} - 2 \Omega g_{t\phi} - \Omega^2 g_{\phi\phi}}}
\end{equation}
where we have neglected the light bending \citep{Lu79,BMT01}.

The flux and the emission spectrum of the accretion disks around compact objects
satisfy some simple scaling relations, with respect to the linear scaling
transformation of the radial coordinate, given by $r\rightarrow \widetilde{r}=r/M$,
where $M$ is the mass of the compact object. Generally, the metric tensor
coefficients are invariant with respect of this transformation, while the
specific energy, the angular momentum and the angular velocity transform as $%
\widetilde{E}\rightarrow \widetilde{E}$, $\widetilde{L}\rightarrow M\widetilde{L}$ and $%
\Omega \rightarrow \widetilde{\Omega}/M$, respectively. The flux scales as $F(r)\rightarrow F(%
\widetilde{r})/M^{4}$, giving the simple transformation law of the temperature as $%
T(r)\rightarrow T\left( \widetilde{r}\right) /M$. By also rescaling the frequency
of the emitted radiation as  $\nu \rightarrow \widetilde{\nu}=\nu /M$,
the luminosity of the disk is given by $L\left( \nu \right) \rightarrow
L\left( \widetilde{\nu}\right) /M$. On the other hand, the flux is proportional
to the accretion rate $\dot{M}_{0}$, and therefore an increase in the
accretion rate leads to a linear increase in the radiation emission flux
from the disk.

The efficiency $\epsilon $ with which the central object converts rest mass into outgoing radiation is the other important physical parameter characterizing the properties of the accretion disks. The efficiency
is defined by the ratio of two rates measured at infinity: the rate of the
radiation of the energy of the photons escaping from the disk surface to infinity, and the rate at which mass-energy is transported to the compact object. If all the emitted photons can escape to infinity, the efficiency depends only on the specific energy measured at the marginally stable orbit $r_{ms}$,
\begin{equation}
\epsilon = 1 - \left.\widetilde{E}\right|_{r=r_{ms}}\;.  \label{epsilon}
\end{equation}

There are some solutions for the geometry of the neutron and quark stars where  $r_{ms}$ is at the surface of the star or even takes values less than the surface radius of the central object. In this case the inner edge of the disk touches the surface of the star and plasma under the effect of any perturbation due to hydro- or magnetohydrodynamic instabilities in the disk will leave the disk and hit the surface. In this case the energy $\widetilde{E}_{e}$ transferred to the star from the disk is measured at the radius $R_e$ of the star, and the efficiency takes the form
$\epsilon = 1 - \widetilde{E}_{e}$, where $\widetilde{E}_{e}=\left.\widetilde{E}\right|_{r=R_e}$. For Schwarzschild black holes the efficiency is about 6\%, no matter if we
consider the photon capture by the black hole, or not. Ignoring the capture
of radiation by the black hole, $\epsilon$ is found to be 42\% for rapidly
rotating black holes, whereas the efficiency is 40\% with photon capture in
the Kerr potential. For neutron and quark stars the efficiency is varying in a broader range, especially if we take into account that $\widetilde{E}_{ms}$ and $\widetilde{E}_{e}$ can have very different values for different neutron and quark star models.

\section{Kerr black holes and CFL quark stars}\label{eos}

In the present Section we briefly review the basic properties of the Kerr black holes and of the CFL quark stars.

\subsection{Kerr black holes}

The Kerr metric, describing a rotating black hole, is given,  in the Boyer-Lindquist coordinate system,  by
\begin{eqnarray}
ds^{2}&=& -\left(1-\frac{2mr}{\Sigma}\right)dt^{2}+2\frac{2mr}
{\Sigma}a\sin^{2}\theta dt d\phi+\frac{\Sigma}{\Delta _K}dr^{2}+
   \nonumber \\
&&\hspace{-0.5cm}\Sigma d\theta^{2}+\left(r^{2}+a^{2}
+\frac{2mr}{\Sigma}a^{2}\sin^{2}\theta\right)\sin^{2}\theta
d\phi^{2},
\end{eqnarray}
where $\Sigma =r^2+a^2\cos ^2\theta$, $\Delta _K=g_{t\phi}^{2}-g_{tt}g_{\phi\phi}=r^2+a^2-2mr$, and $a$ is the Kerr parameter taking values from 0 to $m$. In the equatorial plane, the metric components reduce to
\begin{eqnarray*}
g_{tt} & = &
-\left(1-\frac{2m}{r}\right),\;\;g_{t\phi}  =  2\frac{ma}{r},\\
g_{rr} & = & \frac{r^{2}}{\Delta _K}, \;\; g_{\phi\phi} = r^{2}+a^{2}\left(1+\frac{2m}{r}\right),
\end{eqnarray*}
respectively.

For a Kerr black hole the geodesic equation Eq.~(\ref{geodeqs3}) for $r$ becomes
\begin{equation}
\frac{r^{2}}{\Delta _K}\left(\frac{dr}{d\tau}\right)^{2}=V_{eff}(r),
\end{equation}
with the effective potential given by
\begin{eqnarray}
V_{eff}(r)&=&-1+\nonumber\\
&&\frac{\widetilde{E}^{2}\left[r^{2}(r^{2}+a^{2})
 +2ma^{2}r\right]+4\widetilde{E}\widetilde{L}mar-\widetilde{L}^{2}
 \left(r^{2}-2mr\right)}{
 r^{2}(g_{t\phi}^{2}-g_{tt}g_{\phi\phi})} .
\end{eqnarray}
These relations may be rewritten in the form
\begin{equation}
r^4\left(\frac{dr}{d\tau}\right)^{2}=V(r),
  \label{KerrPot}
\end{equation}
with $V(r)$ given by $V(r)=r^{2}\Delta _KV_{eff}(r)$, respectively.

\subsection{CFL quark stars}

The chemical composition of neutron stars at densities beyond the nuclear saturation remains uncertain, with alternatives ranging from purely nucleonic composition through hyperon or meson condensates to deconfined quark matter \citep{PaRe06}. \citet{Wi84} proposed that strange quark matter (consisting of up $u$, down $d$, and strange $s$ quarks) might be the absolute ground state of hadronic matter at all pressures.  If strange quark matter is the energetically preferred state at very high pressures, but ceases to be energetically preferred below some non-zero threshold minimum pressure, then one would expect the existence of so-called hybrid stars, with strange quark cores, but with exterior regions composed of normal neutron-star matter at pressures below the threshold value.  Therefore the presence of a quark core in a neutron star cannot be ruled out, and the hypothesis of the existence of hybrid stars may be consistent with the high masses and radii indicated by some recent observations. Hybrid stars with quark cores have been intensively investigated in the physical literature \citep{Glen}. In hybrid stars strange quark matter must be in equilibrium with normal nuclear matter. The mixed phase can occur, for a reasonable confinement parameter, near the normal nuclear saturation density, and goes over into pure quark matter at about 5 times the saturation density. The onset of mixed and quark phases is compatible with some observed classes of neutron stars, but it hinders the occurrence of kaon condensation \citep{Glen}. However, when quark seeds form in the core of a neutron star, they will propagate to the entire star and convert it to the new quark phase. A phase transition occurs between the hadronic and quark phase when the pressures and chemical potentials in the two phases are equal, $P_h=P_q$ and $\mu _h=\mu _q$, where $P_h$, $\mu_h$, and $P_q$, $\mu _q$ are the pressures and chemical potentials in the hadronic and quark phase, respectively \citep{Glen}. The change from the metastable  neutron matter phase to the stable quark phase occurs as the result of the thermodynamical fluctuations in a homogeneous medium, formed of neutrons, in which small quantities of the quark phase (called bubbles or nuclei) are randomly generated. Once the quark matter is formed inside the neutron star, it will propagate throughout the entire star. The combustion of the nuclear matter into quark matter can occur either as a slow combustion process, or as a detonation. In both cases, the presence of a quark seed in a neutron star will convert the entire star to a quark star, and the conversion time may be of the order of milliseconds \citep{Chan09}.

In order to obtain a consistent and realistic physical description of the rotating general relativistic quark stars, as a first step we have to adopt the equations of state for the quark matter.
It is generally agreed today that the color-flavor-locked (CFL) state is
likely to be the ground state of matter, at least for asymptotic densities,
and even if the quark masses are unequal \citep{cfl1,cfl2,cfl3}. Moreover,
the equal number of flavors is enforced by symmetry, and electrons are
absent, since the mixture is automatically neutral. By assuming that the
mass $m_{s}$ of the $s$ quark is not large as compared to the chemical
potential $\mu $, the thermodynamical potential of the quark matter in CFL
phase can be approximated as \citep{LuHo02}
\begin{equation}
\Omega _{CFL}=-\frac{3\mu ^{4}}{4\pi ^{2}}+\frac{3m_{s}^{2}}{4\pi ^{2}}-%
\frac{1-12\ln \left( m_{s}/2\mu \right) }{32\pi ^{2}}m_{s}^{4}-\frac{3}{\pi
^{2}}\Delta ^{2}\mu ^{2}+B,
\end{equation}%
where $\Delta $ is the gap energy. With the use of this expression the
pressure $P$ of the quark matter in the CFL phase can be obtained as an
explicit function of the energy density $\varepsilon $ in the form \citep%
{LuHo02}
\begin{equation}\label{pres}
P=\frac{1}{3}\left( \varepsilon -4B\right) +\frac{2\Delta ^{2}\delta ^{2}}{\pi
^{2}}-\frac{m_{s}^{2}\delta ^{2}}{2\pi ^{2}},
\end{equation}
where
\begin{equation}
\delta ^{2}=-\alpha +\sqrt{\alpha ^{2}+\frac{4}{9}\pi ^{2}\left( \varepsilon
-B\right) },
\end{equation}%
and $\alpha =-m_{s}^{2}/6+2\Delta ^{2}/3$. In the following the value of the gap energy $\Delta $ considered in each case will be also mentioned for the CFL equation of state, so that, for example, CFL800 represents the CFL EOS with $\Delta =800$. For the bag constant $B$ we adopt the value $4B=4.2\times 10^{14}$ g/cm$^3$, while for the mass of the strange quark we take the value $m_s=150$ MeV.

For the considered equation of state of quark matter the pressure-density relation is presented  in Fig.~{\ref{fig1}.
\begin{figure}
\centering
\includegraphics[width=8.15cm]{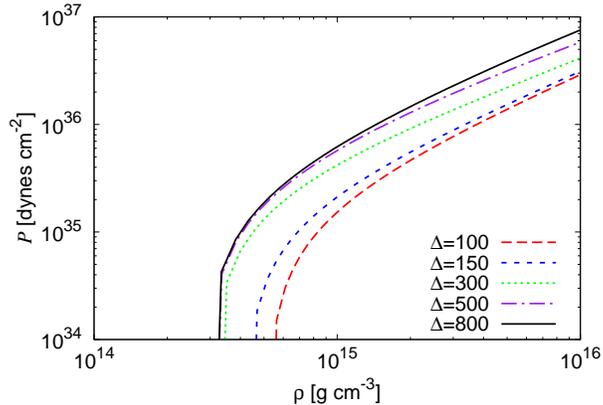}
\caption{Pressure as a function of density (in a logarithmic scale) for EOS CFL for different values of $\Delta$.}
\label{fig1}
\end{figure}

 In order to obtain the emissivity properties of the disk, the metric outside the rotating general relativistic stars must be determined. In the present study we calculate the equilibrium configurations of the rotating quark stars by using the RNS code, as introduced in \citet{SteFr95}, and discussed in detail in \citet{Sterev}. This code was used for the study of different models of rotating neutron stars in \citet{No98} and for the study of the rapidly rotating strange quark stars \citep{Ste99}. The software provides the metric potentials for various types of compact rotating general relativistic objects, which can be used to obtain the physical properties of the accretion disks. As a first step in our study, by using the RNS code we obtain the mass function of rapidly rotating CFL stars,  and determine the maximum mass they can achieve, as well as the physical conditions under which this maximum mass can exist.

The mass distribution of the quark stars in the parameter space
$\left(\rho _c,r_p/r_e\right)$ is represented in Fig.~\ref{fig0}.
In our analysis we consider four values of the gap parameter $\Delta $,
 namely, $\Delta =100, 300, 500$ and 800 MeV, respectively. The central densities corresponding to these stellar configurations are in
 the ranges of $0.52-9\times 10^{15}$ g/cm$^3$ for $\Delta=100$ MeV,  $0.32-9\times 10^{15}$  g/cm$^3$ for $\Delta=300$ MeV, and $0.3-9\times 10^{15}$ g/cm$^3$ for $\Delta=500$ MeV and 800 MeV, respectively. The lower limits of the central densities are determined by the criterion of the positivity of the pressure in the star.
For a given value of $\Delta$ we analyze only the configurations corresponding to the maximum mass.

The physical properties of the models with maximal total mass are presented for rotating CFL stars in Table~\ref{table1}, and in
Table~\ref{table11} for static CFL stars, respectively.
  With the increase of the gap parameter the masses of the rotating CFL quark stars
 increase as well, reaching  a value as high as $5.46M_{\odot}$ for $\Delta =800$ MeV. The increase of the gap parameter $\Delta $
 also produces a considerable increase in the equatorial radii (from 15 km for $\Delta =100$ MeV to 23 km for $\Delta =800$ MeV), and of
 the angular momentum of the rotating CFL quark stars. In each case $R_e>R_{Sch}$, where $R_{Sch}$ is the Schwarzschild radius defined as
 $R_{Sch}=2GM/c^2$. For $\Delta =100$ MeV, $R_{Sch}=7.89$ km, smaller than $R_e=15$ km. For $\Delta =800$ MeV, $R_{Sch}=16.18$ km,
 while $R_e=23$ km.  The angular velocity $\Omega $ decreases from $10^4$1/s to $8\times10^3$1/s with the increasing values of $\Delta$. There is also a moderate change in the position of $h_+$, the height from surface of the last stable co-rotating circular orbit in the equatorial plane. While for $\Delta =100$ MeV, $h_+$ is located on the surface, the last stable co-rotating circular orbits are located at 2 and 3 km from the surface of the star in the equatorial plane for $\Delta =300$ and $\Delta =500$ MeV and $\Delta =800$ MeV, respectively.

 In the case of the static configurations, in Table~\ref{table11} we present the physical characteristics of the CFL stars with the maximum masses corresponding to the given values of $\Delta $. The maximum masses of the CFL quark stars vary from $1.84M_{\odot}$ for $\Delta =100$ MeV to $3.92M_{\odot}$ for $\Delta =800$ MeV, while the stellar radii change from $10$ to $16$ km. The corresponding central densities $\rho _c$ are in the range of $2.4\times 10^(15)$ g/cm$^3$ for $\Delta =100$ MeV and $9\times 10^{14}$ g/cm$^3$ for $\Delta =800$ MeV. Both $h_+$ and $h_-$ have negative values, being located below the surface of the star. The localization of $h_+$ strongly influences the energy flux from accretion disks around compact objects.

For all values of the gap energy $\Delta $ the shape of the curves $M=M\left(r_e\right)$ exhibits the standard form specific to quark stars \citep{Ch98}.
However, the dependence of the mass-radius relation on the central density of the stars is different for the different classes of models. For the low mass - low radius limit of these curves (the left bottom part of the plots), the lowest possible central density (with its value of $5.2\times10^{15}$ g/cm$^3$), is considerably higher for $\Delta=100$ MeV, than the lower limits of $\rho_c$ for $\Delta=300-800$ MeV, located at $3-3.2\times10^{15}$ g/cm$^3$. The $\rho_c$-dependence of the branches
running from the maximal masses down to the lower mass limit for $\Delta=100$,  or for higher values of $\Delta$, is also rather dissimilar. Considering, e.g., the case $r_p/r_e=0.5$, for $\Delta=100$ MeV and for $\rho_c=6\times10^{15}$ g/cm$^3$, we obtain $M=0.70M_{\odot}$ and $R_e=10.72$ km, respectively,  values that are close to the lower limit. The corresponding
values in the mass-radius plane for $\Delta=300$ MeV are $M=4.70 M_{\odot}$ and $R_e=21.49$ km,  and even higher for higher values of $\Delta$, which are located close to the maximum of these curves on the mass scale.  In order to reach the maximal mass for $\Delta=100$ MeV the central density must be set to $2\times10^{15}$ g/cm$^3$. For high values of $\Delta$, the low mass branch of these curves  can therefore be covered by a much more limited range of central densities $\rho_c$, as compared with the case $\Delta=100$ MeV. We conclude that the mass-radius relation for $\Delta\geq300$ MeV is much more sensitive for the change of the central density than it is for $\Delta=100$ MeV. Of course, in the latter case the mass-radius range spanned by the low mass branch is a factor of two smaller in the mass scale than those belonging to  $\Delta\geq300$ MeV,  but the maximal change in $R_e$ is roughly of the order of $10-12$ km for any values of $\Delta$. Therefore for  $\Delta\geq300$ MeV the increase in the equatorial radius is relatively small. As compared to nonrotating stars, the effect of rotation is to increase the equatorial radius of the star and also to increase the mass that can be sustained at a given central energy density.

 One needs to consider stability of the models since models which are unstable with respect to radial (or quasi-radial) pulsational modes are not generally of astrophysical interest. A quark star with low mass is predominantly self-bound, and is stable to these modes. As one moves along the mass-radius curves shown in Fig.~\ref{fig0}, proceeding away from the lowest mass models, the central density progressively increases. In the absence of rotation, the models are stable up to the first maximum in the curve, and unstable thereafter. But with rotation the situation becomes more complicated. If such a sequence of models is calculated at constant angular momentum $J$, then a similar criterion applies \citep{Fried88}, but our sequences are calculated for constant axis-ratio, and so the criterion does not directly apply to them. Roughly speaking, it is likely that stability will continue to change roughly at the maxima, but analysis of this is outside the scope of the present paper.

The differences with respect to the standard quark star models are determined by the change of the sign of the quantity $2\Delta ^2-m_s^2/2$ in the equation of state of the quark matter in CFL phase, Eq.~(\ref{pres}). If $m_s<\sqrt{2}\Delta$, the positive correction term will increase the pressure in the star, by reducing the numerical value of the bag constant, while for $m_s>\sqrt{2}\Delta $, the negative values of the correction term soften the equation of state, and increase the value of the effective bag constant. Hence large $\Delta $ values determine a smaller effective bag constant, which in turn allows larger stellar masses \citep{Wi84}.

\begin{table}
\centering
\begin{tabular}{|l|c|c|c|c|}
\hline
$\Delta \; [{\rm MeV}]$ & 100  & 300 & 500 & 800 \\
\hline
$\rho_c\;[10^{15}{\rm g}/{\rm cm}^{3}]$         & 1.70000 & 0.70000 & 0.61500 & 0.60000 \\
\hline
$M\;[M_{\odot}]$  & 2.67179 & 4.87321 & 5.32598 & 5.46782\\
\hline
$M_0\; [M_{\odot}]$ & 3.20367 & 6.10475 & 6.79655 & 6.96692\\
\hline
$R_e [{\rm km}]$ & 14.8727 & 21.9955 & 22.4372 & 23.2988\\
\hline
$\Omega [10^3{\rm s}^{-1}]$ & 10.7827 & 8.08650 & 8.06520 & 7.89727\\
\hline
$\Omega_p [10^3{\rm s}^{-1}]$  & 10.7541 & 8.02614 & 8.09665 & 7.70780\\
\hline
$T/W [10^{-1}]$ & 2.08674 & 2.34702 & 2.22506 & 2.36814 \\
\hline
$cJ/GM_{\odot}^2$ & 6.24670 & 21.1302 & 24.8740 & 26.5042 \\
\hline
$I [10^{45}{\rm g}{\rm cm}^2]$ & 5.09138 & 22.9645 & 27.1046 & 29.4951 \\
\hline
$\Phi_2 [10^{45}{\rm g}{\rm cm}^2]$ &  0.87825 & 3.95150 & 4.50791 & 5.10189 \\
\hline
$h_+ [{\rm km}]$ & $-$  & 2.83756 & 3.27146 & 3.02038 \\
\hline
$h_- [{\rm km}]$ & - & - & - & - \\
\hline
$\omega_c/\Omega [10^{-1}]$ & 7.58826 & 8.51031 & 8.89081 & 8.80415 \\
\hline
$r_e  [{\rm km}]$ & 9.95880 & 12.4564 & 11.7922 & 12.4806\\
\hline
$r_p/r_e$ & 0.47000 & 0.47000 & 0.50000 & 0.47000 \\
\hline
\end{tabular}
\caption{Physical parameters of rotating maximum mass CFL quark stars for different values of $\Delta$. Here $\rho _c$ is the central density, $M$ is the gravitational mass, $M_0$ is the rest mass, $R_e$ is the circumferential radius at the equator, $\Omega $ is the angular velocity, $\Omega _p$ is the angular velocity of a particle in circular orbit at the equator, $T/W$ is the rotational-gravitational energy ratio, $cJ/GM_{\odot}^2$ is the angular momentum $I$ is the moment of inertia, $\Phi_2$ is the mass-quadrupole moment, $h_+$ is the height from surface of the last stable co-rotating circular orbit in the equatorial plane, $h_{-}$ is the height from surface of the last stable counter-rotating circular orbit in the equatorial plane, $\omega _c/\Omega$ is the ratio of the central value of the potential $\omega $ to $\Omega $, $r_e$ is the coordinate equatorial radius and $r_p/r_e$ is the axes ratio (polar to equatorial).}
\label{table1}
\end{table}

\begin{table}
\centering
\begin{tabular}{|l|l|l|l|l|}
\hline
$\Delta$ [MeV] & 100 & 300 & 500 & 800\\
\hline
$\rho_c\;[10^{15}{\rm g}/{\rm cm}^{3}]$ & 2.4 & 1 & 0.9 & 0.9 \\
\hline
$M$ $[M_\odot]$ & 1.84841 & 3.41121 & 3.84898 & 3.92723 \\
\hline
$M_0$ $[M_\odot]$ & 2.24186 & 4.39966 & 5.08115 & 5.2071\\
\hline
$R_e$ [km] & 9.88262 & 15.1998 & 16.173 & 16.2828\\
\hline
$\Omega$ $[10^4 s^{-1}]$ & 0 & 0 & 0 & 0 \\
\hline
$\Omega_p$ $[10^4 s^{-1}]$ & 1.5859 & 1.12729 & 1.09043 & 1.0899 \\
\hline
$T/W$ & 0 & 0 & 0 & 0 \\
\hline
$cJ/GM_\odot^2$ & 0 & 0 & 0 & 0\\
\hline
$I [{\rm g}{\rm cm}^2]$ & - & - & - & -\\
\hline
$\Phi_2 [{\rm g}{\rm cm}^2]$ & 0 & 0 & 0 & 0\\
\hline
$h_+$ [km] & 0 & 0 & -0.329715 & -4.42561\\
\hline
$h_-$ [km] & 0 & 0 & -0.329715 & -0.442561\\
\hline
$\omega_c/\Omega$  & 0 & 0 & 0 & 0\\
\hline
$r_e$  [km] & 6.88092 & 9.51013 & 9.67292 & 9.63205\\
\hline
$r_p/r_e$ & 1 & 1 & 1 & 1\\
\hline
\end{tabular}
\caption{Physical parameters of maximum mass static CFL quark stars.}
\label{table11}
\end{table}

\begin{figure}
\centering
\includegraphics[width=.48\textwidth]{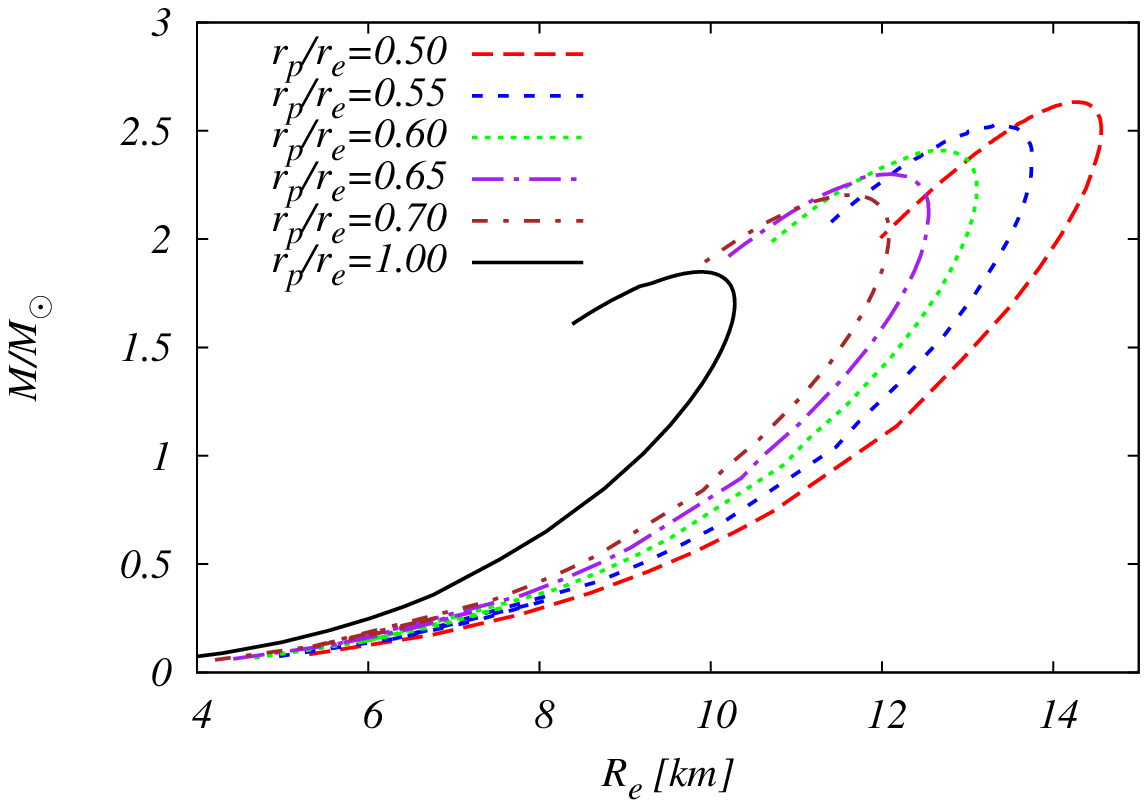}
\includegraphics[width=.48\textwidth]{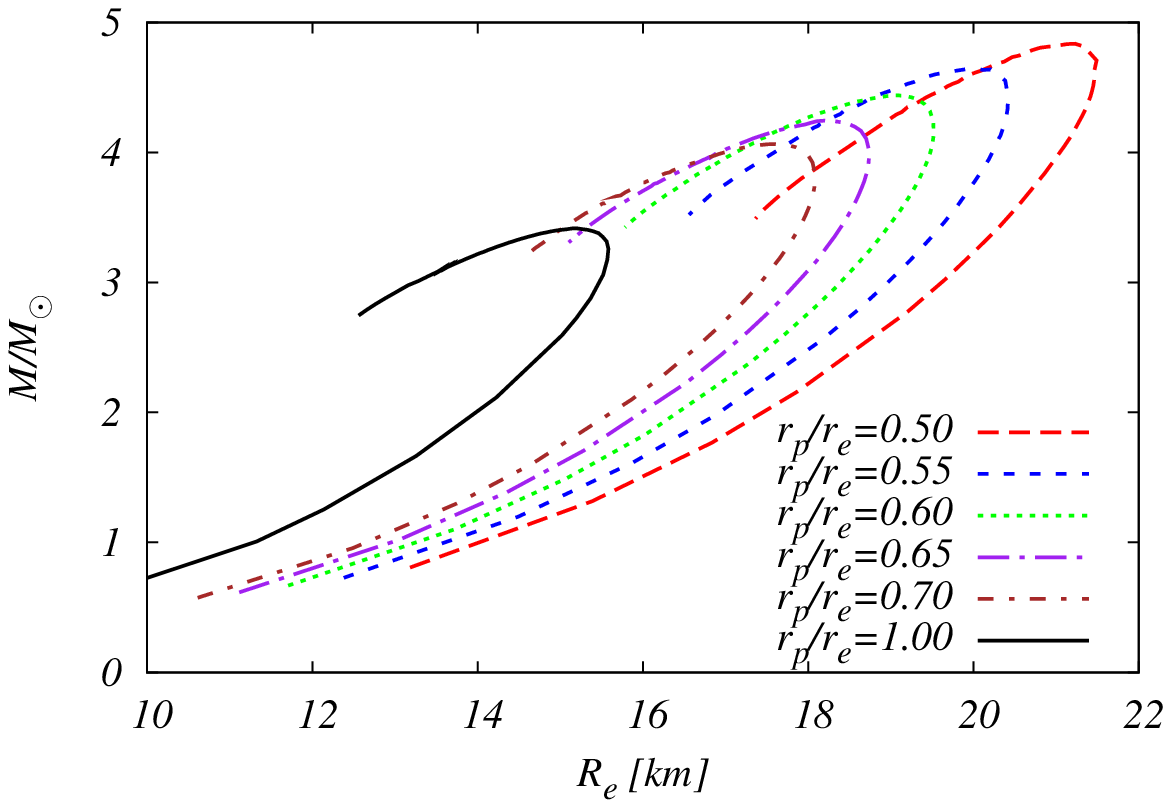}\\
\includegraphics[width=.48\textwidth]{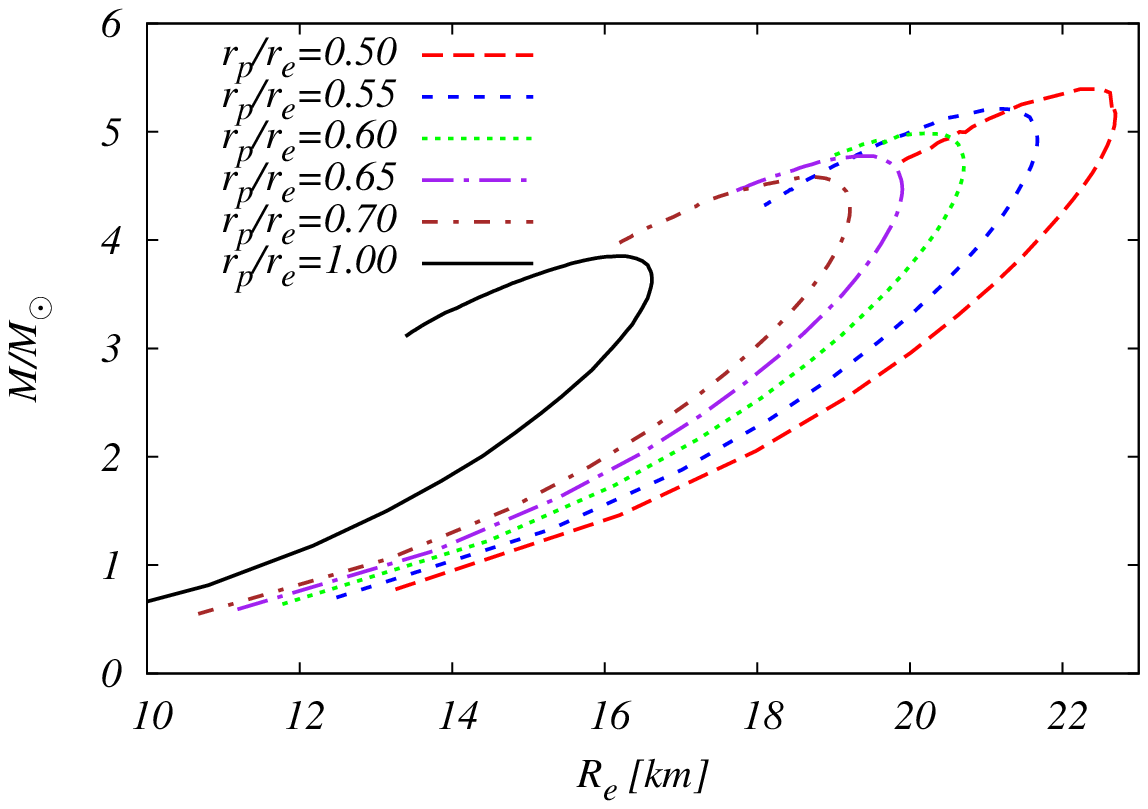}
\includegraphics[width=.48\textwidth]{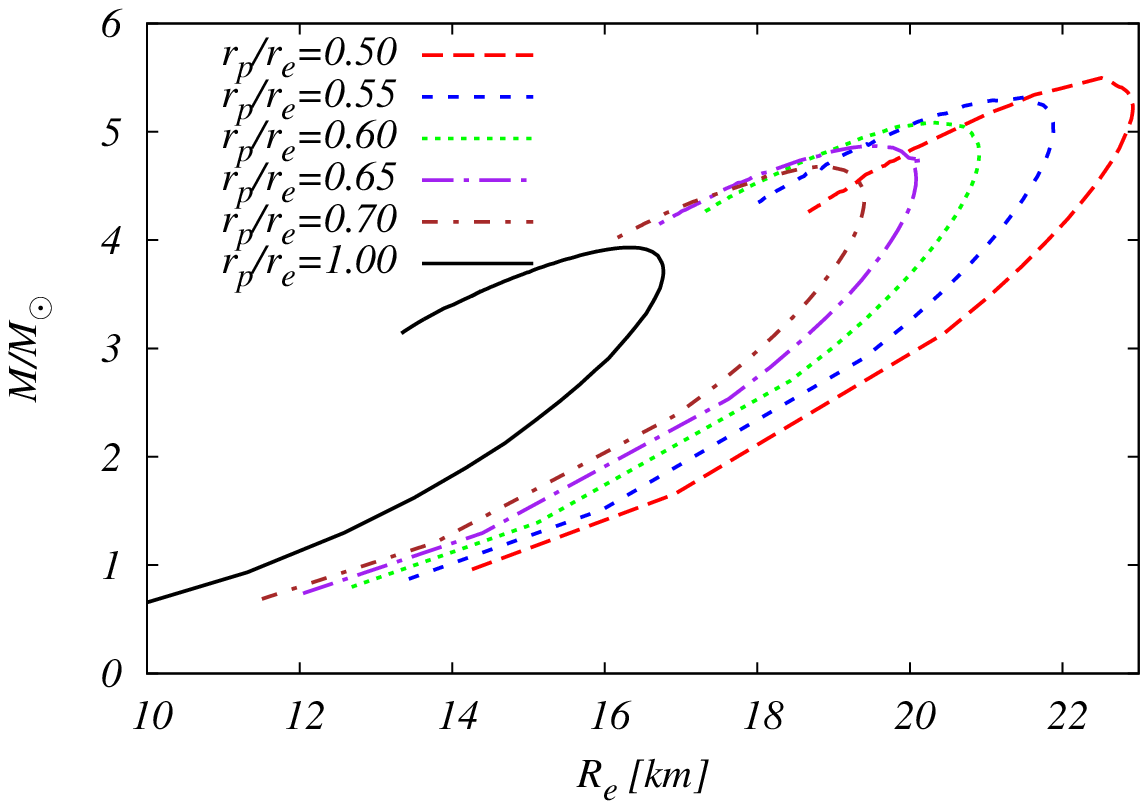}
\caption{Gravitational mass versus equatorial circumferential radius of the rapidly rotating and static CFL quark stars for different values of the gap energy $\Delta $: $\Delta =100$ MeV (upper left figure), $\Delta =300$ MeV (upper right figure), $\Delta =500$ MeV (lower left figure), and $\Delta =800$ MeV (lower right figure).}
\label{fig0}
\end{figure}

\section{Disk accretion onto Kerr black holes and quark stars}\label{41}

The metric outside the rotating compact general relativistic stars can be described, in quasi-isotropic coordinates, as
\begin{equation}
ds^2=-e^{\bar{\gamma }+\bar{\rho }}dt^2+e^{2\bar{\alpha } }\left(d\bar{r}^2+\bar{r}^2d\theta ^2\right)+
e^{\bar{\gamma }-\bar{\rho }}\bar{r}^2\sin ^2\theta \left(d\phi -\bar{\omega}dt\right)^2,
\end{equation}
where the metric potentials $\bar{\gamma }$, $\bar{\rho }$, $\bar{\alpha }$ and the angular velocity of the stellar fluid relative to the local inertial frame  $\bar{\omega}$ are all functions of the quasi-isotropic radial coordinate $\bar{r}$ and of the polar angle $\theta $. The RNS code produces the metric functions in a quasi-spheroidal coordinate system, as functions of the parameter $s=\bar{r}/\left(\bar{r}+\bar{r}_e\right)$, where $\bar{r}_e$ is the equatorial radius of the star, which we have converted into Schwarzschild-type coordinates $r$ according to the equation $r=\bar{r}\exp\left[\left(\bar{\gamma }-\bar{\rho }\right)/2\right]$. To obtain the radius of the innermost stable circular orbits $r_{ms}$ we have used a truncated form of the analytical approximation given by \citet{ShSa98},
\begin{eqnarray}
r_{ms}&\approx &6M\left(1-0.54433q-0.22619q^2+0.17989Q_2-0.23002q^2+\right.\nonumber\\
&& \left. 0.26296qQ_2-0.29693q^4+0.44546q^2Q_2\right),\
\end{eqnarray}
where $q=J/M^2$ and $Q_2=-M_2/M^3$, respectively, and where $J$ is the spin angular momentum, and $M_2$ is the quadrupole moment.

Once the metric outside the rotating stars is known, from Eqs.~(\ref{tildeL})-(\ref{Omega}) we obtain the angular velocity, the specific energy and the specific angular momentum of the orbiting plasma particles in the disk. Then we calculate the integral given by Eq.~(\ref{F}), measuring the photon flux emitted by the disk surface in thermodynamical equilibrium. From the flux we obtain the temperature distribution of the disk, as well as the spectra of the emitted radiation.  For all our calculations we use an accretion rate of $\dot M=1\times10^{-12} M_{\odot}$/yr, unless specified otherwise. In order to compare the thermal and emission properties of the thin accretion disks onto black holes and quark stars, we consider several classes of models in which both the Kerr black hole and the quark stars have the same mass and angular velocity. The physical parameters of the rotating CFL quark stars used in the present study are presented in Table~\ref{table3}. The static case, with lower maximal mass than in the rotating case, is also plotted for comparison.

\begin{table}
\centering
\begin{tabular}{|l|c|c|c|c|}
\hline
$\Delta \; [{\rm MeV}]$ & 100  & 300 & 500 & 800 \\
\hline
$\rho_c\;[10^{15}{\rm g}/{\rm cm}^{3}]$         & 1.50000 & 1.75000 & 0.61500 & 0.60000 \\
\hline
$M\;[M_{\odot}]$  & 2.01099 & 3.80791 & 5.04850 & 5.46782\\
\hline
$M_0\; [M_{\odot}]$ & 2.40633 & 4.76377 & 6.43507 & 6.96692\\
\hline
$R_e [{\rm km}]$ & 11.4128 & 16.1807 & 21.1469 & 23.2988\\
\hline
$\Omega [10^3{\rm s}^{-1}]$ & 7.70653 & 7.71710 & 7.69576 & 7.89727\\
\hline
$\Omega_p [10^4{\rm s}^{-1}]$  & 1.33266 & 1.06480 & 8.61193 & 7.70780\\
\hline
$T/W [10^{-2}]$ & 8.29980 & 11.5253 & 20.6856 & 23.6814 \\
\hline
$cJ/GM_{\odot}^2$ & 2.22209 & 9.45230 & 21.3040 & 26.5042 \\
\hline
$I [10^{45}{\rm g}{\rm cm}^2]$ & 2.53405 & 10.7646 & 24.3289 & 29.4951 \\
\hline
$\Phi_2 [10^{45}{\rm g}{\rm cm}^2]$ & 2.13271 & 10.2406 & 37.3391 & 51.0189 \\
\hline
$h_+ [{\rm km}]$ & 0.00000 & $-$ & $-$ & 3.02038 \\
\hline
$h_- [{\rm km}]$ & $-$ & -0.155693 & 0.00859 & 4.78999 \\
\hline
$\omega_c/\Omega [10^{-1}]$ & 6.44529 & 8.76271 & 8.61004 & 8.80415 \\
\hline
$r_e  [{\rm km}]$ & 7.99607 & 9.19234 & 11.1606 & 12.4806\\
\hline
$r_p/r_e$ & 0.78000 & 0.73500 & 0.57000 & 0.47000 \\
\hline
\end{tabular}
\caption{Physical parameters of the rotating CFL quark stars used for the calculation of the electromagnetic properties of the thin accretion disks.}
\label{table3}
\end{table}

In Figs.~\ref{fig2} we present the radiated flux distribution over the accretion disk around rotating black holes and quark stars, for different values of the total mass $M$ and of the spin parameter $a_*=J/M^2$ of the central compact objects. The four plots show that the inner edge of the accretion disk is located at a lower radius,  and the amplitude of the flux radiated from the disk surface increases with increasing total mass and spin parameter for both type of objects.
However, the inner radius $r_{in}$ is greater, and, consequently, the values of $F(r)$ are smaller for quark stars than for black holes in each configuration of $M$ and $a_*$. For comparison we also plotted the flux profiles of the accretion disk for non-rotating CFL stars with the maximum equilibrium mass $M_{static}$. Since the geometry outside the static star is given by the Schwarzschild metric, the properties of the flux profiles for the static star coincide with those for static black holes. The inner edge of the disk is located at the radius $r=6M$, and the maximal flux is lower than the one of the rotating cases, in spite of the fact that $M_{static}$ is smaller than $M$. The latter indicates that rotation has a considerable effect in producing a higher disk radiation flux, as compared to the disk luminosities for the static cases.
All these distinctive features for the rotating and static cases can be seen in the profiles of the disk temperature as well, which are presented in Figs.~\ref{fig3}.

\begin{figure}
\centering
\includegraphics[width=.48\textwidth]{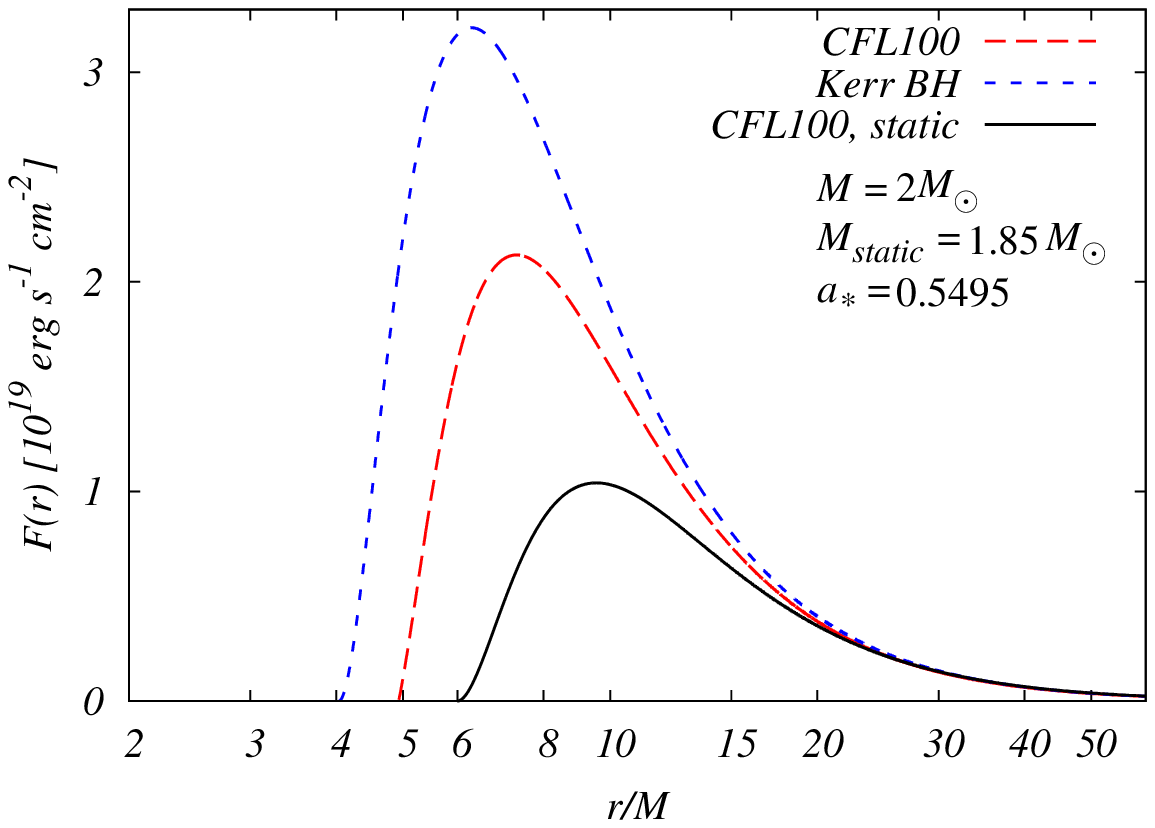}
\includegraphics[width=.48\textwidth]{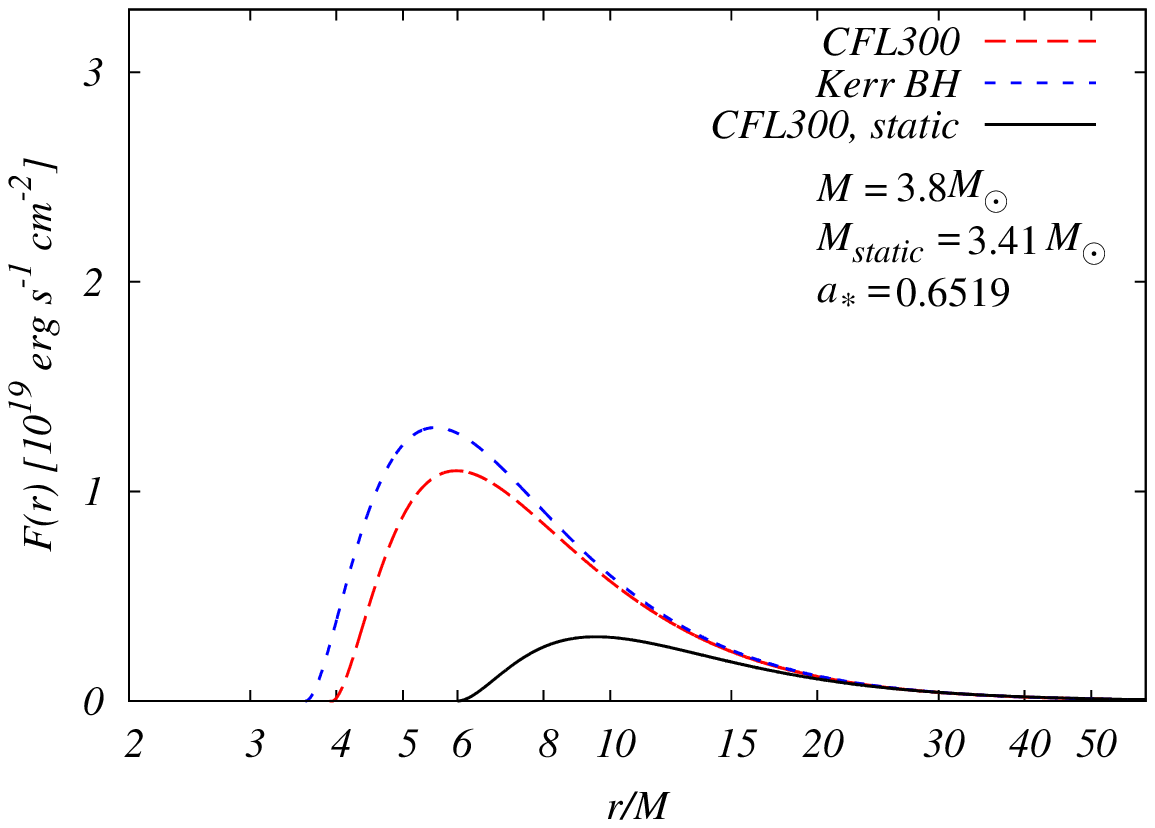}\\
\includegraphics[width=.48\textwidth]{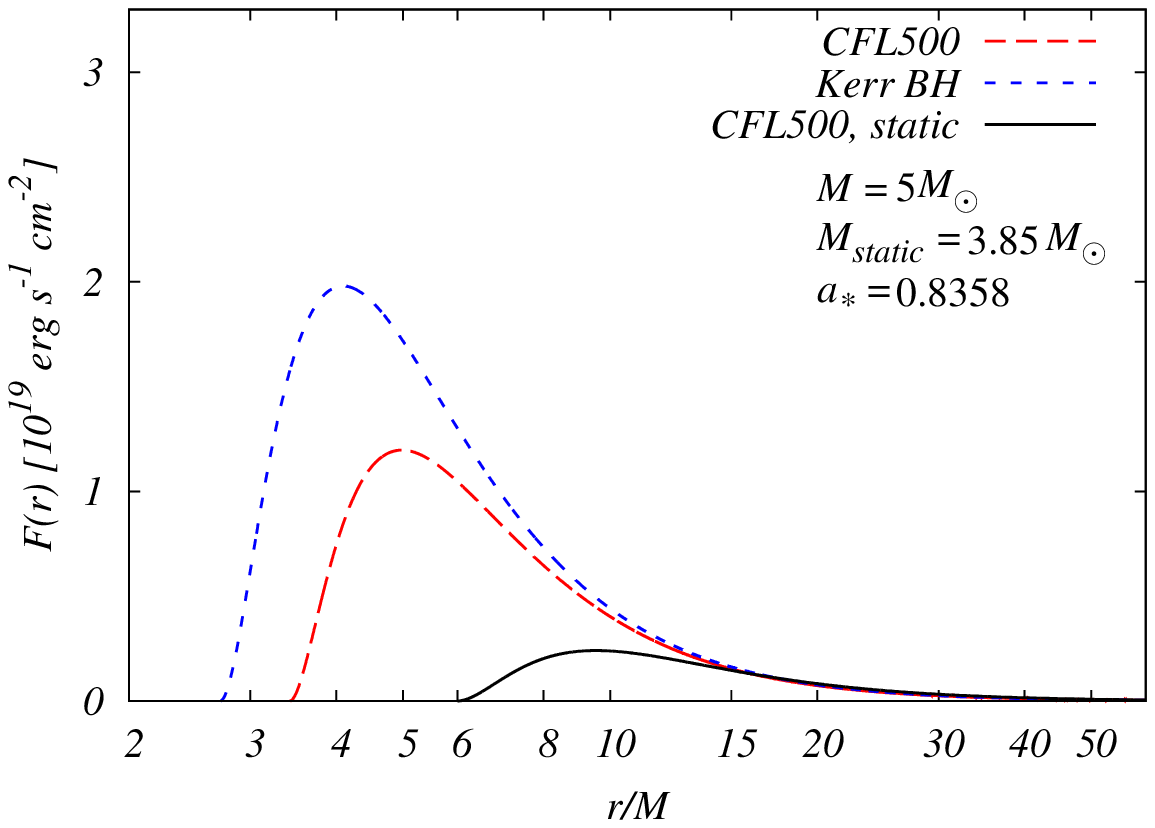}
\includegraphics[width=.48\textwidth]{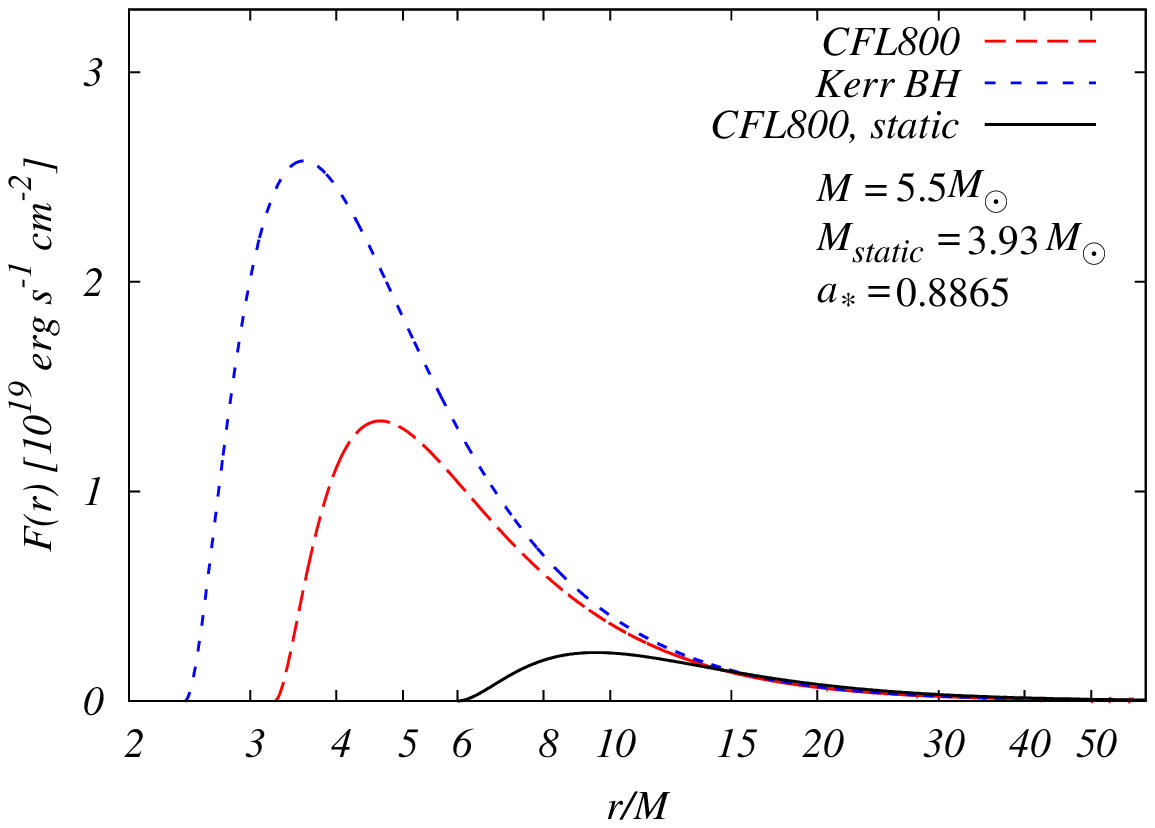}
\caption{The time-averaged flux radiated by a thin accretion disk around rotating black holes and quark stars with the same total mass $M$ and spin parameter $a_{*}$. The static case with lower maximal mass is also plotted for comparison. }
\label{fig2}
\end{figure}

\begin{figure}
\centering
\includegraphics[width=.48\textwidth]{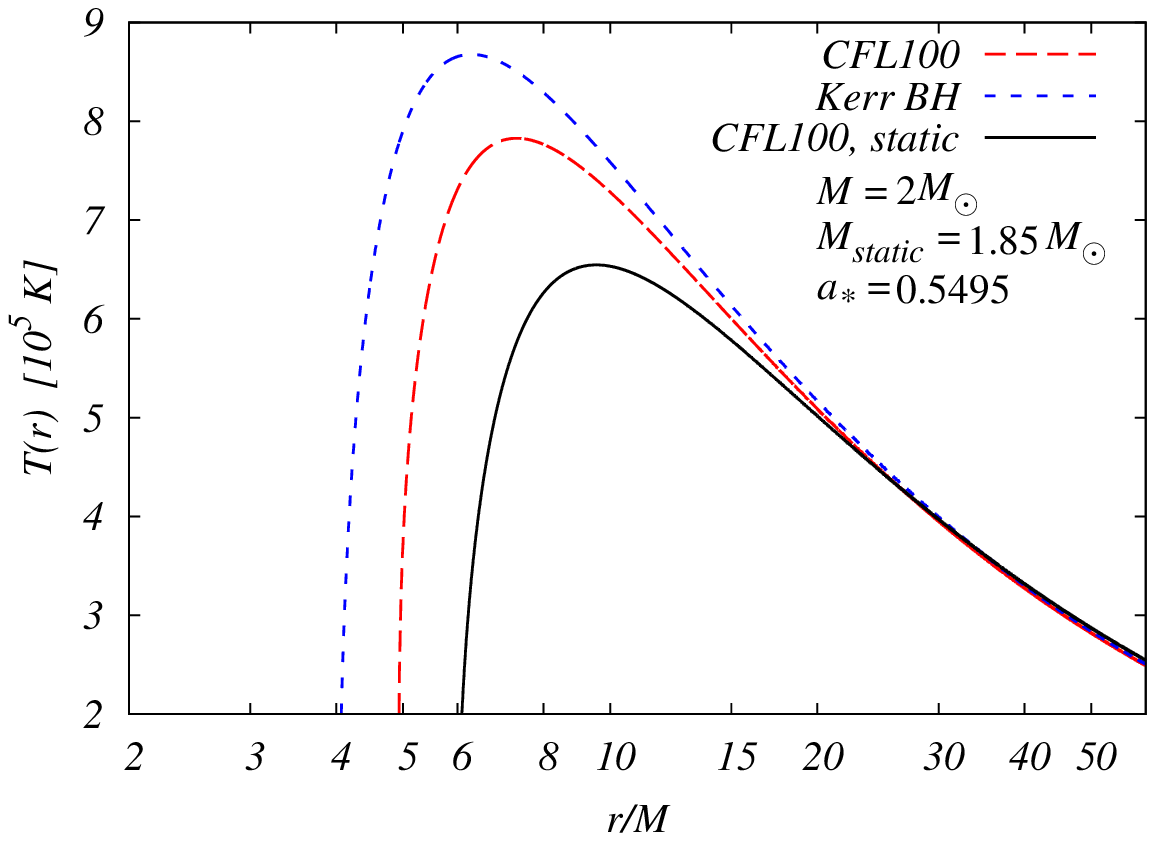}
\includegraphics[width=.48\textwidth]{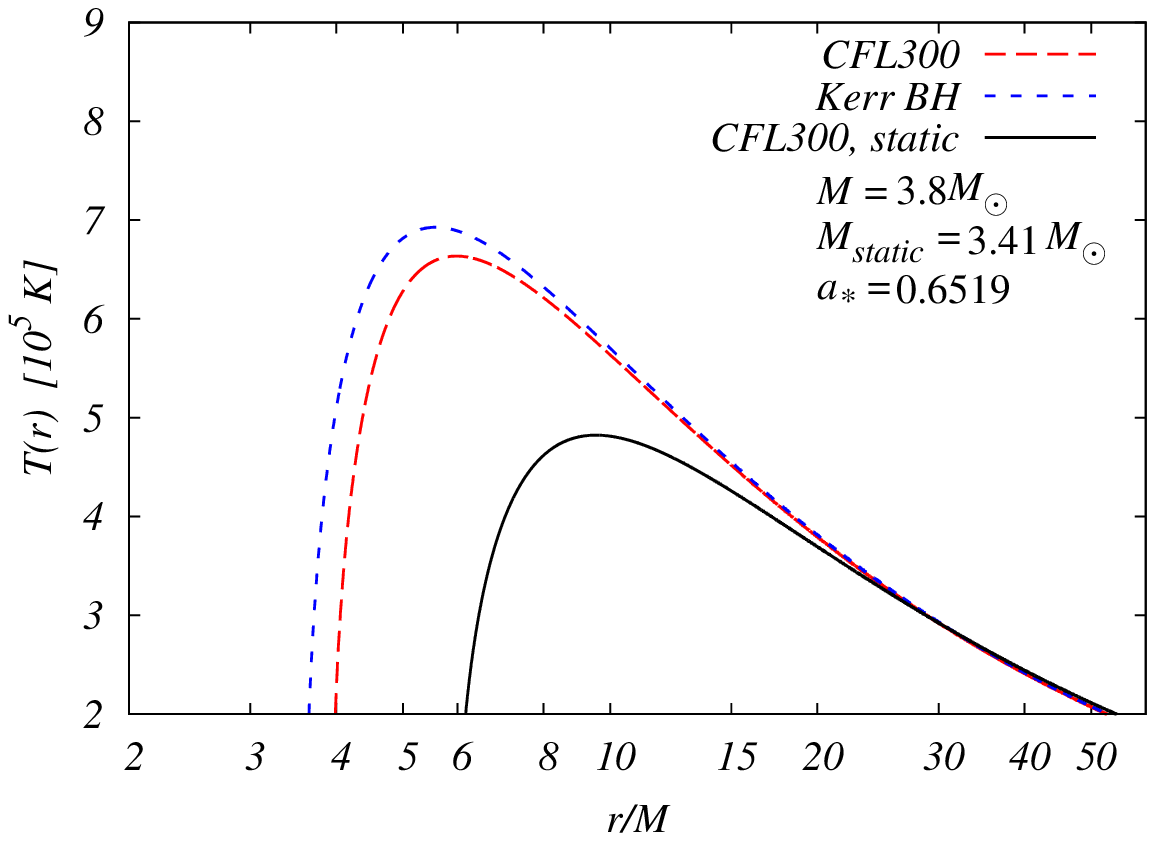}\\
\includegraphics[width=.48\textwidth]{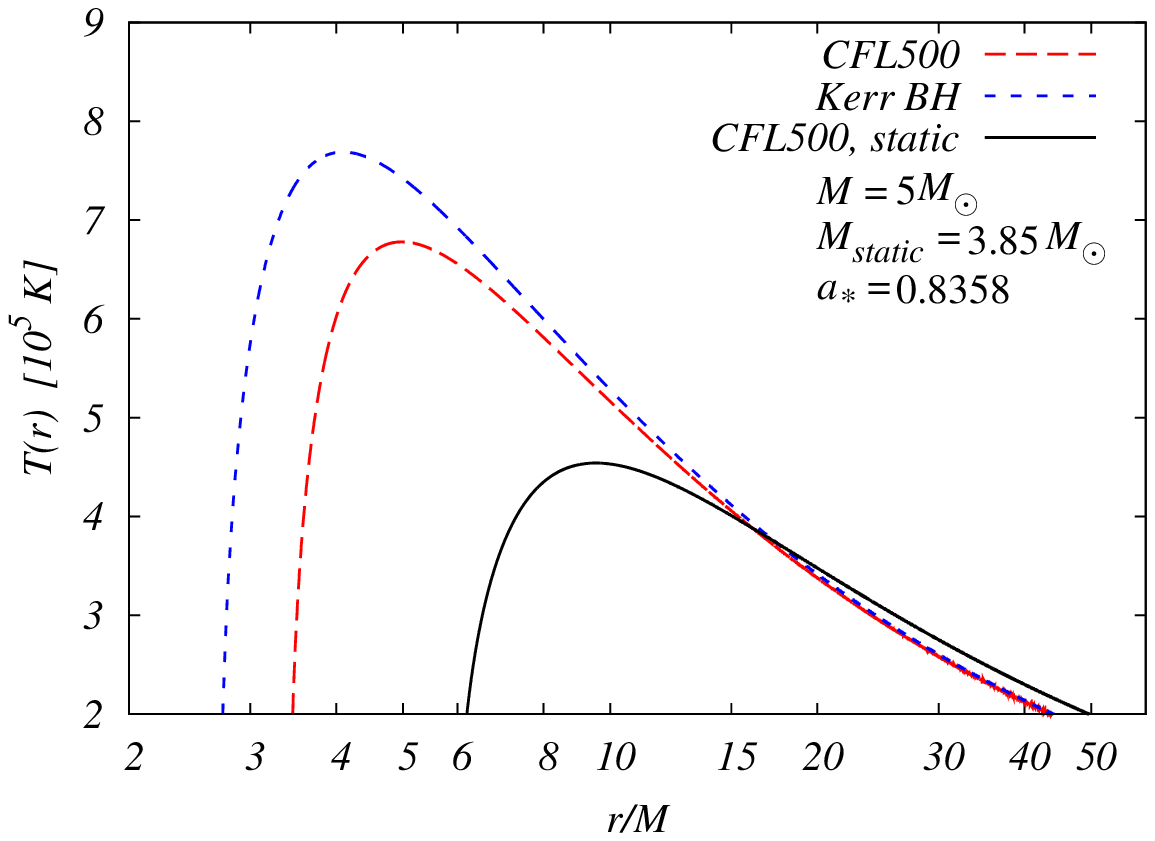}
\includegraphics[width=.48\textwidth]{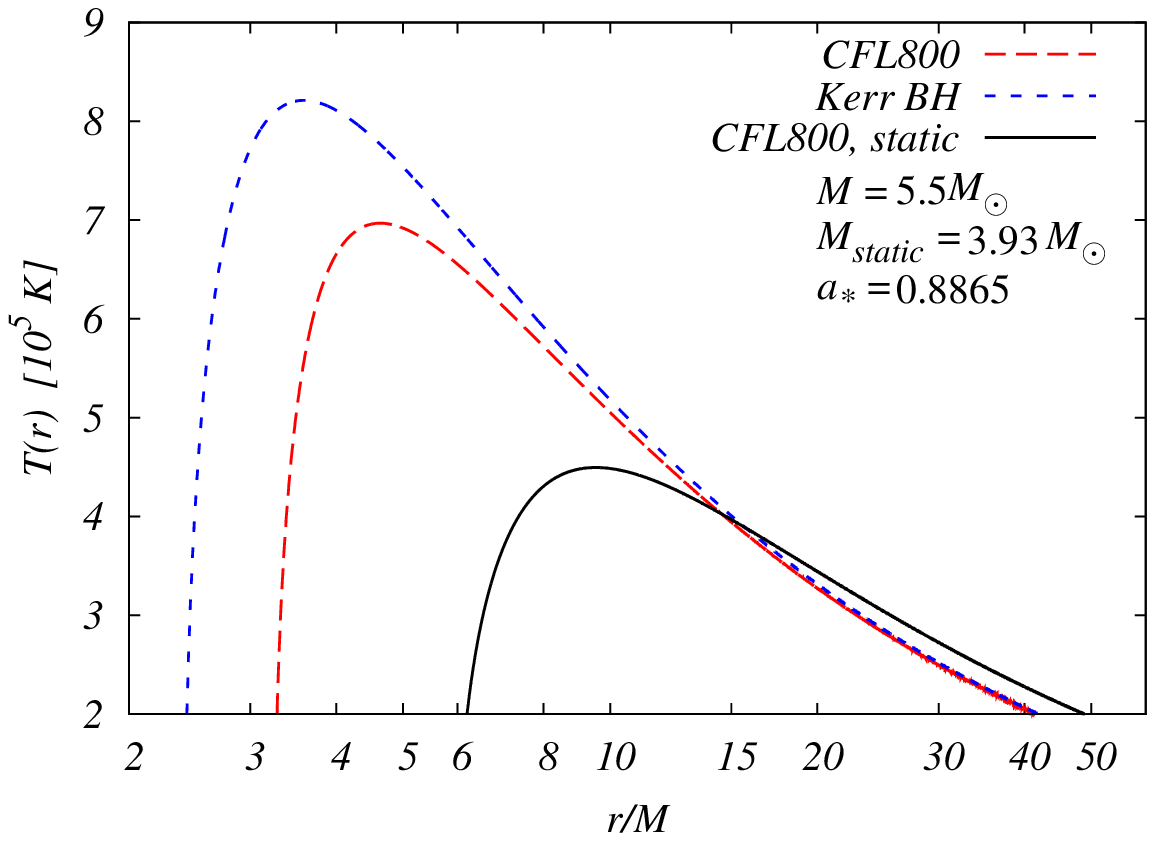}
\caption{The temperature profiles of the thin accretion disk around rotating black holes and quark stars with the same total mass $M$ and spin parameter $a_{*}$. The static configuration is also presented.}
\label{fig3}
\end{figure}

The characteristics of the disk spectra around black holes and quark stars also exhibit significant differences. For black holes, the maxima of the spectra are located at higher frequencies, and reach higher values than those for quark stars.
In case of the static CFL quark stars with the maximal total mass, both the spectral amplitude, and the frequency at which the maximum is located, have lower values than those for the rotating cases (with higher maximal masses).

\begin{figure}
\centering
\includegraphics[width=.48\textwidth]{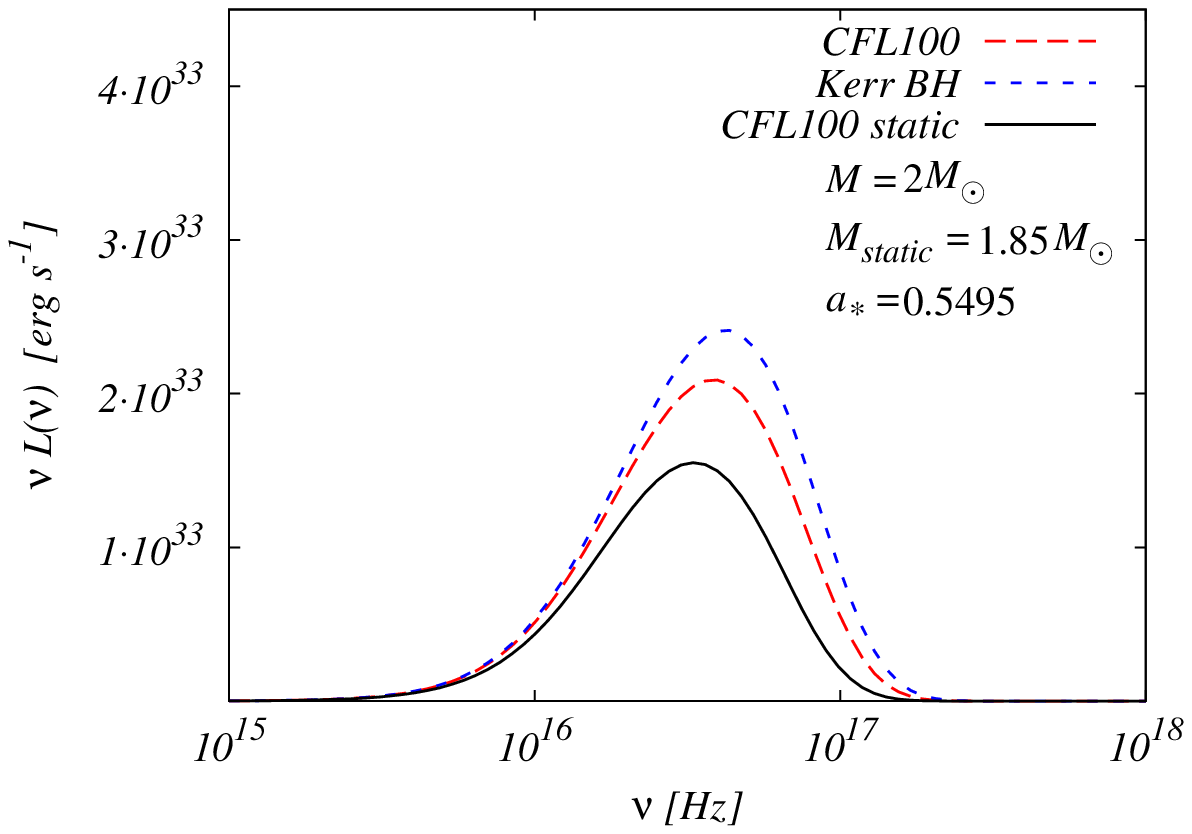}
\includegraphics[width=.48\textwidth]{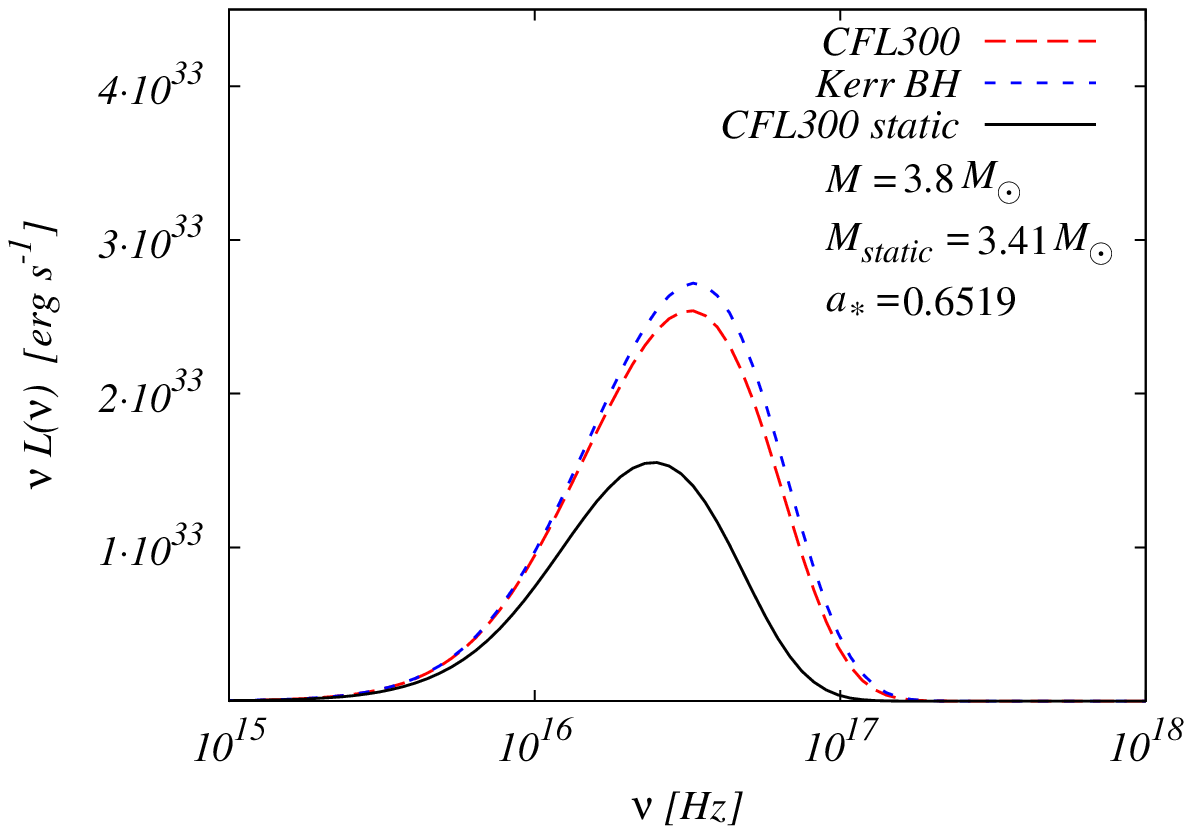}\\
\includegraphics[width=.48\textwidth]{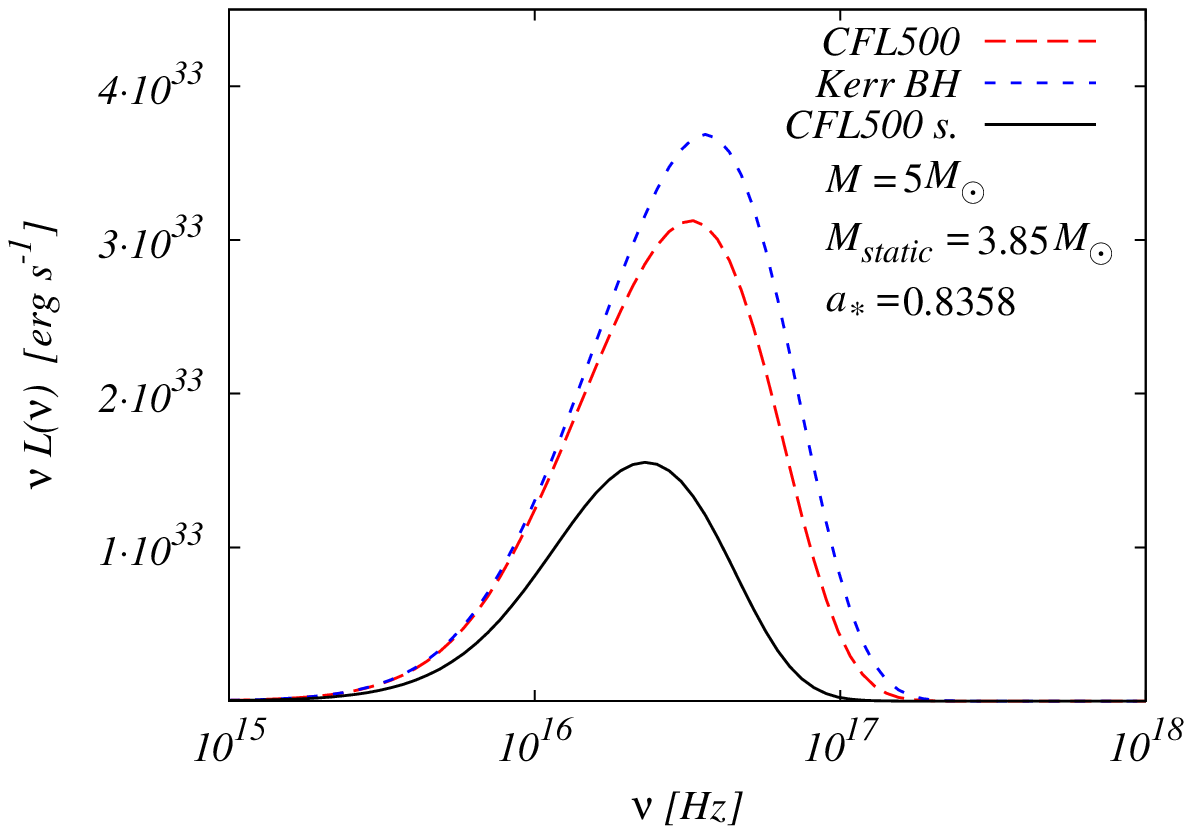}
\includegraphics[width=.48\textwidth]{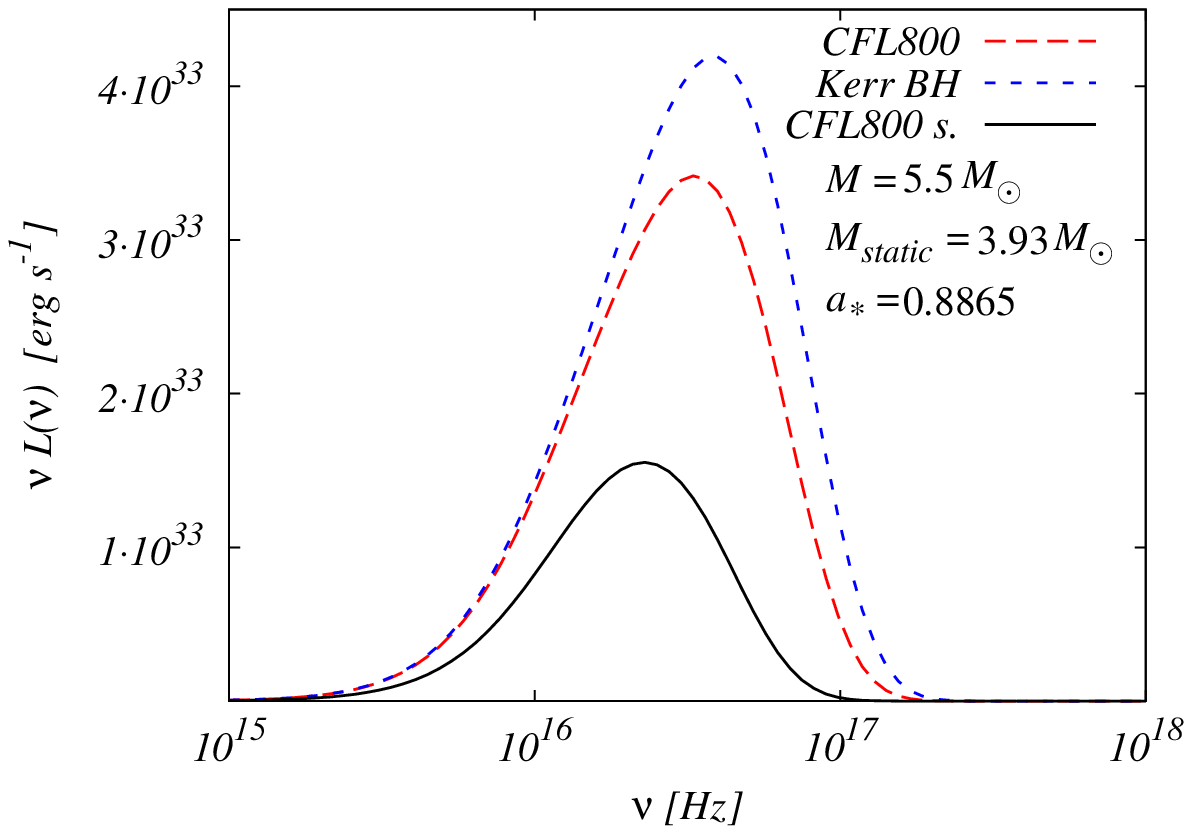}
\caption{The disk spectra for rotating black holes and quark stars with the same total mass $M$ and spin parameter $a_{*}$. The static case }
\label{fig4}
\end{figure}

\section{Discussions and final remarks}

Table~\ref{table2} shows the conversion efficiency $\epsilon$ of the accreted mass into radiation for both rotating black holes and quark stars. For a given configuration of total mass $M$ and spin parameter $a_*$, $\epsilon$ is somewhat higher in the accretion process driven by black holes, as compared to the values obtained for rotating  CFL quark stars. This means that rotating black holes can always convert more efficiently mass into radiation than CFL type rotating quark stars can do.

\begin{table}
\begin{center}
\begin{tabular}{|c|c|c|c|c|c|}
\hline
& $M$ $[M_{\odot}]$ & 2 & 3.8 & 5 & 5.5\\
\hline
& $a_*$  & 0.5494. & 0.6519 & 0.8358 & 0.8865 \\
\hline
EOS CFL & $r_{in}$ [km] & 14.64  & 21.97  &  24.84 & 26.31 \\
        & $\epsilon$   & 0.0762 & 0.0921 & 0.1130 & 0.1228 \\
\hline
BH &$r_{in}$ [km]  & 11.99  & 20.30  & 20.25 & 19.46\\
    & $\epsilon$  & 0.0863 & 0.0971 & 0.1315 & 0.1495\\
\hline
\end{tabular}
\end{center}
\caption{The radius of the inner disk edge $r_{in}$ and the efficiency $\epsilon $ for quark stars with EOS CFL and Kerr black holes with the same total mass and spin parameter.}
\label{table2}
\end{table}

As shown by the flux integral in Eq.~(\ref{F}), and the explicit expressions of the specific energy, specific angular momentum and angular velocity given by Eqs.~(\ref{tildeE}) - (\ref{Omega}),  the rather different characteristics of the radial flux distribution over the accretion disk, of the disk spectra, and of the conversion efficiency, respectively, are due to the differences between the metric potentials of the black holes and of the quark stars, respectively. Even if the total mass and the angular velocity are the same for each type of the rotating central object, producing similar radial profiles for $\Omega$, $\widetilde E$ and $\widetilde L$, the radiation properties of the accretion disks around these objects exhibit large differences. The reason is that the proper volume, and in turn the function $\sqrt{-g}$, used in the calculation of the flux integral, is highly dependent on the behavior of the metric component $g_{rr}=(\partial \overline{r}/\partial r)^2 g_{\overline{r}\overline{r}}$. The latter contains the $r$-derivatives of the metric functions $\rho(r)$ and $\gamma(r)$, respectively, via the coordinate transformation between $\overline{r}$ and $r$, which are extremely sensitive to the slopes of $\rho(r)$ and of $\gamma(r)$, respectively. As a result, though the inner edges of the disks are located at almost the same radii in the flux and temperature plots presented in Section~\ref{41}, the amplitudes of the energy fluxes emerging from the disk surfaces around black holes and CFL quark stars are significantly different.  Similar differences also appear in the disk spectra of quark stars and black holes, respectively.

The Coulomb barrier at the quark surface of a hot strange quark star may also be a
powerful source of $e^{+}e^{-}$ pairs, which are created in the extremely
strong electric field of the barrier. At surface temperatures of around $%
10^{11}$ K, the luminosity of the outflowing plasma may be of the order $%
\sim 10^{51}$ ergs$^{-1}$ \citep{Us98a, Us98b,HaCh06}. Moreover, as shown by
\citet{PaUs02}, for about one day for normal quark matter and for up to a
hundred years for superconducting quark matter, the thermal luminosity from
the star surface, due to both photon emission and $e^{+}e^{-}$ pair
production may be orders of magnitude higher than the Eddington limit.

Photon emissivity is the basic parameter for determining macroscopic
properties of stellar type objects.  \citet{Al86}
have shown that, because of very high plasma frequency $\omega _p$ near the strange quark
matter edge, photon emissivity of strange quark matter is very low. Propagation of electromagnetic waves of
frequencies lower than $\omega _{p}$ is exponentially damped. Therefore, only photons
produced just below the surface with momenta pointing outwards can leave
strange quark matter. For temperatures $T<<E_{p}/\omega $, where $E_{p}\approx 23$
MeV is the characteristic transverse plasmon cutoff energy, the equilibrium
photon emissivity of strange quark matter is negligible small, compared to the
black body one. The spectrum of equilibrium photons is very hard, with $%
\hbar \omega >20$ MeV. The problem of the soft photon emissivity of quark matter at the surface
of strange quark stars has been considered in \citet{ChHa03} and \citet{HaCh05}. By taking into account the
Landau-Pomeranchuk-Migdal effect and the absorption of the radiation in the
external electron layer, the emissivity of the quark matter can be six
orders of magnitude lower than the equilibrium black body radiation. However, despite these very specific signatures for quark stars, a definite method for discriminating them with respect to the neutron stars is still missing.  Quark stars have a very low photon emissivity, which make them very difficult to directly identify them observationally.

Unpaired strange quark matter in  bulk contains $u$, $d$, $s$ quarks as well as electrons to maintain charge neutrality. The chemical balance is maintained by weak interactions and neutrinos assumed to escape from system. If strange quark matter is in the CFL phase, in which quarks of all flavors and colors near the Fermi surface form pairs, an equal number of flavors is enforced by symmetry, and the mixture is neutral \citep{cfl3}. Therefore CFL quark stars do not have an electrosphere, and the main energy loss mechanism is through quark-quark bremsstrahlung \citep{ChHa03, Chan09}. The luminosity $L_S$ of the accreting quark star is given by $L_S=\eta \dot{M}c^2$, where for the energy conversion rate $\eta $ we can take the value $\eta \approx 0.1$. For an accretion rate $\dot{M}=10^{-12}M_{\odot}/{\rm year}$, the luminosity of the star is $L_S\approx 5.78\times 10^{33}\;{\rm erg/s}$, while the luminosity $L_d$ of the disk for the same accretion rate is, depending on the value of the gap energy $\Delta $, in the range $L_d\approx 2\times 10^{33}-4\times 10^{33}\;{\rm erg/s}$. Therefore the luminosities of the disk and of the quark star are very close, having approximately the same values. On the other hand, the surface temperature of the central accreting quark star is given by $T_S=\left(\eta \dot{M}c^2/4\pi R^2\epsilon \sigma \right)^{1/4}$, where $R$ is the radius of the star, and $\epsilon =10^{-4}-10^{-5}$ \citep{ChHa03, Chan09}. For $\epsilon =10^{-5}$, the surface temperature $T_S$ of the star is of the order of $T_S\approx 3\times 10^7$ K, while the corresponding temperature of the disk $T_d$ is around $T_d\approx 6\times 10^5-9\times 10^5$ K. Therefore the temperature of the star is much larger than that of the disk, $T_S>>T_d$. Furthermore, while the spectral distribution $L_d\left(\nu \right)$ of the radiation coming from the disk is a standard black body distribution, the surface radiation of the quark star has a typical bremsstrahlung spectrum \citep{HaCh05}. Therefore, the observational detection of a bremsstrahlung type electromagnetic emission from the central object, in combination with the emission properties of the accretion disk, could provide a distinct signature for the presence of a high mass and extremely low luminosity quark star in a black hole candidate system.

In conclusion, the observational study of the thin accretion disks around rapidly rotating compact objects can provide a powerful tool in distinguishing between standard black holes and stars with exotic equations of state, that have underwent, for example, a phase transition from the neutron phase to a quark phase, as well as for discriminating between the different phases of the dense matter.

\section*{Acknowledgments}

We would like to thank to the anonymous referee for comments and suggestions that helped us to significantly improve the manuscript. K. S. C. is supported by the GRF grant number HKU 7013/06P of the government of the Hong Kong SAR. The work of T. H. is supported by the GRF grant number 702507P of the Government of the Hong Kong SAR. Z. K. is indebted to the colleagues in the Department of Physics and Center for Theoretical and Computational Physics of the University of Hong Kong for their support and warm hospitality.

\end{document}